\documentclass[
  journal=pasa,
  manuscript=research-paper,
  year=2023,
  volume=37,
]{cup-journal}

 \usepackage[utf8]{inputenc}

\usepackage{graphicx}

\usepackage{amsmath}
\usepackage[nopatch]{microtype}
\usepackage{booktabs}

\usepackage{hyperref}

\definecolor{darkblue}{rgb}{0.0, 0.0, 0.5}
\definecolor{darkred}{rgb}{0.8, 0.0, 0.0}

\hypersetup{colorlinks=true,citecolor=darkblue,linkcolor=darkblue}

\makeatletter
\DeclareRobustCommand{\HI}{%
  \mbox{H\check@mathfonts\fontsize\sf@size\z@\selectfont I}%
}
\makeatother
\DeclareUnicodeCharacter{02BC}{{'}}

\title{WALLABY Pilot Survey: kNN identification of perturbed galaxies through \HI{}{} morphometrics}

\author{Benne W. Holwerda}
\affiliation{University of Louisville, Department of Physics and Astronomy, 102 Natural Science Building, 40292 KY Louisville, USA.}
\email[B.W. Holwerda]{benne.holwerda@louisville.edu}

\author{Helga D\'{e}nes}
\affiliation{School of Physical Sciences and Nanotechnology, Yachay Tech University, Hacienda San Jos\'{e} S/N, 100119, Urcuqu\'{i}, Ecuador }

\author{Jonghwan Rhee}
\affiliation{International Centre for Radio Astronomy Research (ICRAR), University of Western Australia, 35 Stirling Hwy, Crawley, WA 6009, Australia}

\author{Denis Leahy}
\affiliation{University of Calgary, Department of Physics and Astronomy, 2500 University Dr. NW, Calgary, AB, Canada, T2N 1N4}

\author{B\"{a}rbel S. Koribalski}
\affiliation{Australia Telescope National Facility, CSIRO, Space and Astronomy, P.O. Box 76, NSW 1710, Australia}
\alsoaffiliation{School of Science, Western Sydney University, Locked Bag 1797, Penrith, NSW 2751, Australia}

\author{Niankun Yu}
\affiliation{National Astronomical Observatories, Chinese Academy of Sciences, Beijing 100101, Peopleʼs Republic of China}
\alsoaffiliation{Key Laboratory of Radio Astronomy and Technology, Chinese Academy of Sciences, Beijing 100101, Peopleʼs Republic of China}

\author{Nathan Deg}
\affiliation{Department of Physics, Engineering Physics, and Astronomy, Queen’s University, Kingston, ON, K7L 3N6, Canada}

\author{Tobias Westmeier}
\affiliation{International Centre for Radio Astronomy Research (ICRAR), University of Western Australia, 35 Stirling Hwy, Crawley, WA 6009, Australia}

\author{Karen Lee-Waddell}
\affiliation{Australian SKA Regional Centre}

\author{Yago Ascasibar}
\affiliation{Departamento de F\'{i}sica Te\'{o}rica, Universidad Aut\'{o}noma de 
Madrid (UAM), Madrid 28049, Spain}
\alsoaffiliation{Centro de Investigaci\'{o}n Avanzada en F\'{i}sica Fundamental 
(CIAFF-UAM), Madrid 28049, Spain}

\author{Manasvee Saraf}
\affiliation{International Centre for Radio Astronomy Research (ICRAR), University of Western Australia, 35 Stirling Hwy, Crawley, WA 6009, Australia}

\alsoaffiliation{Australia Telescope National Facility, CSIRO, Space and Astronomy, P.O. Box 1130, Bentley, WA 6102, Australia}
\alsoaffiliation{ARC Centre of Excellence for All-Sky Astrophysics in 3 Dimensions (ASTRO 3D), Australia}

\author{Xuchen Lin} % [0000-0002-4250-2709]
\affiliation{Department of Astronomy, School of Physics, Peking University, Beijing 100871, People's Republic of China}

\author{Barbara Catinella}
\affiliation{International Centre for Radio Astronomy Research, The University of Western Australia, Crawley, WA 6009, Australia}
\alsoaffiliation{ARC Centre of Excellence for All-Sky Astrophysics in 3 Dimensions (ASTRO 3D), Australia}

\author{Kelley Hess}
\affiliation{Chalmers University of Technology, Onsala Space Observatory, G\"{o}teborg, Sweden}


\begin{document}

\begin{abstract}
Galaxy morphology in stellar light can be described by a series of ``non-parametric'' or ``morphometric'' 
parameters, such as concentration-asymmetry-smoothness, Gini, $M_{20}$, and Sersic 
fit. These parameters can be applied to column density maps of atomic hydrogen (\HI). The \HI{} distribution is susceptible to perturbations by environmental effects, e.g. inter-galactic medium pressure and tidal interactions. Therefore, \HI{} morphology can potentially identify galaxies undergoing ram-pressure stripping or tidal interactions. We explore three fields in the WALLABY Pilot \HI{} survey and identify perturbed galaxies based on a k-nearest Neighbor (kNN) algorithm using an \HI{} morphometric feature space. For training, we used labeled galaxies in the combined 
NGC 4808 and NGC 4636 fields with six \HI{} morphometrics to train and test a kNN classifier. The 
kNN classification is proficient in classifying perturbed galaxies with all metrics --accuracy, precision 
and recall-- at 70-80\%. By using the kNN method to identify perturbed galaxies in the deployment 
field, the NGC 5044 mosaic, we find that in most regards, the scaling relations of perturbed and 
unperturbed galaxies have similar distribution in the scaling relations of stellar mass vs star 
formation rate and the Baryonic Tully-Fisher relation, but the \HI{} and stellar mass relation flatter 
than of the unperturbed galaxies. Our results for NGC 5044 provide a prediction for future studies 
on the fraction of galaxies undergoing interaction in this catalogue and to build a training sample to 
classify such galaxies in the full WALLABY survey.
\end{abstract}

\section{Introduction}

The atomic gas (\HI) disk extends well beyond the stellar disk of spiral galaxies at the same surface density \cite[e.g.,][for examples and discussions on \HI\ disks]{Bosma78,Begeman89, Meurer96, Meurer98, Swaters02, Noordermeer05, Walter08, Boomsma08, Elson11, Heald11a, Heald11b, Zschaechner11b, de-Blok08, Koribalski18, de-Blok20}. For comparison, see \cite{Trujillo20,Chamba22} for a discussion on the defined edge of stellar disks. The outer regions of these disks are sensitive to ram-pressure stripping by the inter-galactic medium \citep[IGM,][]{Wang21,Reynolds21,Reynolds22}. A lopsided  appearance of the outer \HI\ disk \citep{Jog09,van-Eymeren11a,van-Eymeren11b,Koribalski18} or an asymmetry \citep{Giese16, Reynolds20} can be attributed to tidal interactions \citep{Jog09, Koribalski09}, ram-pressure stripping \citep{Moore98,Westmeier13,Hess22}, a lopsided dark matter halo \citep{Jog02}, ongoing mergers, or a combination of these. The \HI{} in the outer part of disk galaxies is much more sensitive to gravitational (tidal) interaction as well as pressure interactions with the group or cluster medium than the stellar component \citep[e.g.,][]{Hibbard01}. As galaxies are pre-processed in groups,one of the first signs of tidal interactions will be the changes in their gas disks. It is likely that the \HI{} asymmetry is caused by either tidal interaction, or ram-pressure stripping, or both \citep{Yu22d,Watts21}. But some internal perturbation could affect the \HI{} distribution in a similar way, such as AGN feedback \citep[e.g.,][]{Villaescusa-Navarro16,Morganti17} or stellar feedback \citep[e.g.,][]{Ashley17}, or mergers \citep{Zuo22}. As with warps in the \HI{} disk or stellar disk truncations, multiple mechanisms, both internal and external, could be responsible.

Parameterisation of \HI{}{} disk appearance is different from stellar parametrization because the \HI{}{} disk is based on line emission and therefore has a much lower dynamic range: high density \HI{} would become molecular hydrogen while low density \HI{} is difficult to detect and lack of self-shielding would result in transition to ionized hydrogen. The area covered by an \HI{} disk is larger, but the spatial resolution is typically an order of magnitude lower due to the much larger \HI{} beam (or PSF) compared to optical imaging. The morphometric parameter space is one used extensively in ultraviolet/optical images of galaxies; the C-A-S \citep{CAS}, Gini-$M_{20}$ \citep{Lotz04}, DIM \citep{Rodriguez-Gomez19} and S\'{e}rsic profile \citep{Sersic68}.

Here, we apply the galaxy morphometrics originally developed for stellar disks which were applied with some success on \HI{} data in the past \citep{Holwerda11a,Holwerda11b,Holwerda11c,Holwerda11d,Holwerda11e,Holwerda12c,Giese16,Reynolds20,Reynolds23,Deg23,Holwerda23} but on often heterogeneous data. For example \cite{Giese16} pointed out that these \HI{} morphometrics depend strongly on the signal-to-noise ratio of each object, complicating their use across surveys or with varying s/n. \cite{Reynolds20} illustrated the challenge to compare morphometrics, specifically asymmetry, across different \HI{} surveys. The optimal application is therefore within a single survey and a well-documented implementation, i.e. {\sc statmorph} implementation of these morphometrics \citep{Rodriguez-Gomez19}. 
Here we use {\sc statmorph}, a python based tool to compute the most commonly used galaxy morphometrics on already segmented images. This tool is public and uses the commonly used definitions of each morphology parameter and fits a single S\'{e}rsic profile to the light distribution. It was developed for ultraviolet/optical/near-infrared imaging but translates well to \HI{} images \citep{Holwerda23}.\\

% I go through the first three sentences mention "STATMORPH", then I think it is still not clear for the reader what "STATMORPH" is. It seems that it is the main tool used in this paper, I would suggest that you use 2-3 sentence summarize what it is, what's the main function of it, how to understand the derived results.

\HI{} morphometrics are a potential feature space for machine learning algorithms. One could classify if galaxies are undergoing ram-pressure stripping, tidal interactions, or even ongoing mergers based on their position in the \HI{} morphology space. The caveat is that a sufficient training set has to be available. Ideally, the training set spans the input feature space and all the possible use cases. 
Our goal here is to examine how well one can train a simple classifier based on the \HI{} catalogue of a single field of galaxies observed by Widefield ASKAP L-band Legacy All-sky Blind surveY \citep[WALLABY,][]{Koribalski12,Koribalski20}.
% update K12 in bib file: https://ui.adsabs.harvard.edu/abs/2012PASA...29..359K/abstract
and generalize the results to other groups. The \HI{} morphometric space is familiar but it remains unclear which morphometrics are most useful to identify \HI{} perturbations. Our goals break down into how well one can get an interaction fraction in a given group, i.e. a population characteristic and how well one can identify individual galaxies as undergoing a disturbance, be it tidal or ram-pressure stripping. 

The WALLABY \citep{Koribalski12, Koribalski20} is an interferometric \HI{} survey carried out with the Australian Square Kilometer Array Pathfinder \citep[ASKAP,][]{ASKAP,Hotan21}, which provides an ideal laboratory for \HI{} morphometrics. The survey is of uniform image quality and will cover a large fraction of the sky and the local Universe. In the future the WALLABY pipeline will create higher resolution postage stamps for pre-selected galaxies but we use the present pipeline products here. 

Deep, high-resolution, and uniform \HI{} maps, across S/N, resolution, and sensitivity, for a large number of galaxies allows us to compare across environments. The WALLABY pilot survey \citep{Westmeier22,Kim23a,Courtois23,Grundy23} has observed several groups and clusters of galaxies. Here, we use the data of three fields centered on groups of galaxies: NGC 4636, NGC 4808, and NGC 5044 to analyze the effects of environmental effects on \HI{} morphology. One of these groups, NGC 4636, has been examined in detail by \cite{Lin23d} and assessed for signs of ram-pressure stripping, tidal effects and mergers. Their labels extend into the NGC 4808 field as well. 
We will use labeling in these fields as our training/testing sample and the sources in the remaining mosaic on NGC 5044 as the application sample. 

Throughout we use the Planck (2015) cosmology  \citep[$\mathrm{H_0} = 67.74~  \mathrm{km /s / Mpc},~ \Omega_0 = 0.3075$,][]{Planck-Collaboration15}. We adopt a Chabrier initial mass function \citep[IMF,][]{Chabrier03} to uniformly derive SFRs and stellar masses.
The paper is organized as follows: 
section \ref{s:data} describes briefly the WALLABY pilot survey and other data products used, 
section \ref{s:morphometrics} details the definitions of the morphometric parameters used, 
section \ref{s:ML} introduces the machine learning algorithm used and the input considerations, 
section \ref{s:results} shows the results of the kNN classification effort, 
section \ref{s:discussion} discusses these results in context of future uses, and 
section \ref{s:conclusions} are our conclusions. \\

\begin{table*}
    \centering
    \begin{tabular}{l | lllllll}
Field   & RA        & DEC   & Group Distance    & cz        & No. ASKAP fields  & No. Objects   & \\
        & (deg)     & (deg) & (Mpc)             & km/s      & \#                & Full SoFiA catalog            & Resolved ($D<60$ Mpc)\\
\hline
\hline
Virgo   &           &       &           &           &                   & & \\
NGC 4636 & 190.7084 & 2.6880 & 16.2     & 919       & 1                 & 231 & 48\\
NGC 4808 & 193.953958 & 4.304111 & 16.0 & 760       & 1                 & 147 & 89\\
\hline
NGC 5044 & 198.849875 & -16.385528 & 45.7 &         & 4                 & 1326 & 258\\
    \end{tabular}
    \caption{Basic properties of the three galaxy group WALLABY fields analysed here.}
    \label{t:wallaby:data}
\end{table*}

\section{WALLABY Data}
\label{s:data}

The WALLABY survey \citep{Koribalski12,Koribalski20} is an all-sky H\,{\sc i} survey carried out with wide-field Phased Array Feeds (PAFs) on the Australian Square Kilometre Array Pathfinder \citep[ASKAP,][]{ASKAP, Hotan21}. ASKAP consists of $36 \times 12$-m telescopes forming a 6-km diameter interferometer. The PAFs are used to from 36 overlapping beams and together deliver a field-of-view of $\sim$30 square degrees with a resolution of 30$''$ and 4~km\,s$^{-1}$. Before the start of full survey operations, a number of fields were observed in early science and pilot survey programs 
\citep{Serra15a, Lee-Waddell19, Kleiner19, For19, For21}.

The Phase~2 WALLABY pilot survey is described in detail in \cite{Westmeier22}. The pilot data was made available to the collaboration for initial science projects.
This includes the single tile on NGC 4808, NGC 4636, and Vela fields and the 4-tile mosaic in the direction of the NGC 5044 group. Thanks to improvements in data quality and source finding with SoFiA \citep{Serra15b, Westmeier21}, 
the total number of \HI{} detections is higher in the final pilot data. The WALLABY data of these three groups are described in detail in Murugeshan et al. (2024, \textit{submitted}). The ASKAP interferometric WALLABY survey has a beam size of $\sim$30" and an rms of 1.6~mJy\,beam$^{-1}$ for a velocity resolution of 4 km/s \citep{Koribalski20}.

\subsection{Virgo (NGC 4636 and NGC 4808 Fields)}
\label{ss:virgo}

WALLABY's Phase~2 pilot program observed two close fields, each centered on one of two  Virgo groups, NGC 4636 \citep{Lin23d} and NGC 4808 (Murugeshan et al. 2024, {\em submitted}). NGC 4636 is a relatively close group at a distance of 16.2 Mpc \citep{Kourkchi17} and a radius of 0.61 Mpc, based on ROSAT X-ray measurements  \citep{Reiprich02}. Two galaxies, NGC 6156 (in the Norma Field) and NGC 4632 (in this field) were studied in detail in \cite{Deg23} as they show a polar ring structure in \HI. \cite{Lin23d} presents a catalog of galaxies around the N4636 group center with redshift measurements from several \HI{} and optical catalogues. \cite{Lin23d} note that of the 19 galaxies detected by WALLABY belonging to this group, six galaxies are resolved enough for detailed moment-0 map study. They present flags for different types of interaction based on the combined WALLABY-FAST data, which include objects in the NGC 4808 field. This is the basis for our training sample (see section 5.1).
The second WALLABY pilot field is centered around NGC 4808 group. The Tully-Fisher (T-F) distances in this field are presented in \cite{Courtois23}. This group is similarly close, at approximately $\sim$16 Mpc. 

\subsection{NGC 5044 Mosaic} 
\label{ss:n5044}

The third field, a mosaic of four fields centered on the NGC 5044 group, has been studied across wavelengths before in the optical, x-ray and \HI{} observations \citep{Ferguson90, Ferguson91,Tamura03, Buote03, Buote04, Osmond04, McKay04, Forbes06}. 

\begin{figure}
    \centering
    \includegraphics[width=\textwidth]{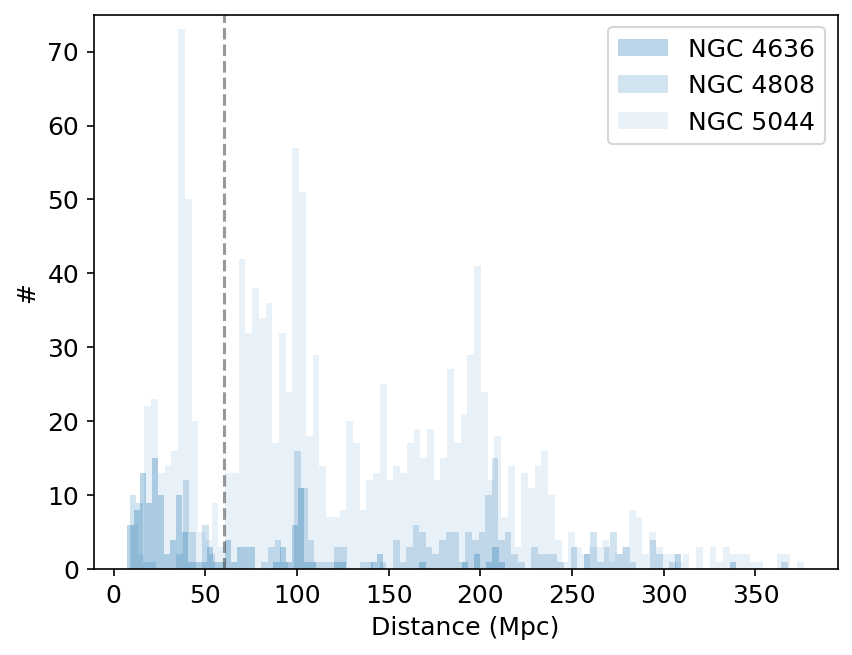}
    \caption{Distribution of distance for galaxies in the three WALLABY fields, centered on NGC 4808, NGC 4646, and NGC 5044. The vertical dashed line is the 60 Mpc cutoff for selection for the training sample in NGC 4808 and the application samples. }
    \label{f:distances}
\end{figure}

The WALLABY internal release on this field (DR3) covers 120 deg$^2$ of the NGC 5044 four-tile mosaic across a 21 cm \HI{} line red shift range of $cz \sim 500$ to 25,400 km/s ($z < 0.085$), which uses the full RFI-free bandwidth available to WALLABY. NGC 5044 DR3 includes 1326 detections. The resulting catalogue is richest of the three in source counts with a large number of sources well behind the nearby group around NGC 5044 (Figure \ref{f:distances}).

Figure \ref{f:distances} shows the distribution of \HI{} detections in all three catalogs. A majority of sources in the three fields is \textit{not} associated with the groups themselves. In the NGC 5044 field especially, several groups and clusters can be identified in the background. There are full samples of these fields for which we compute \HI{} morphometrics and assign stellar mass and star-formation rates.

\section{Stellar Mass and Star-Formation Rates}
\label{s:mstar}

To ensure uniform stellar mass and star-formation rates over the three data-sets, we adopt the WISE photometry derived stellar mass and star-formation rates expressions as described in \cite{Jarrett11,Jarrett13,Cluver14}. For stellar mass, we use their eq. 2 with an absolute solar magnitude $M_{W1} = 3.24$ (W1, Vega), and for the SFR their equation 5,  with $M_{W3} = 3.24$ (W3, Vega). 
Solar luminosities are from \cite{Willmer18} and for each object, we search the ALLWISE catalog accessible through IPAC. Stellar mass-to-light ratios are based on the {\sc w1mpro} -- {\sc w2mpro} colour, the stellar mass is derived from the above mass-to-light ratio, the {\sc w1mpro} and the distance derived from the \HI{} redshift (Figure \ref{f:distances}). Star-formation is based on the {\sc w3mpro} for all galaxies and the equation 5 in \cite{Cluver14}. Like all single color-based mass-to-light ratios and single band star-formation indicators, these estimates are approximate with a greater degree of uncertainty than dedicated Spectral Energy Distribution modeling results \citep[cf][]{de-los-Reyes24}. In a similar vein, the distance derived from the \HI{} redshift may be influenced by peculiar motion within the group. We opted for the \HI{}-redshift derived distances and the WISE derived stellar mass and star-formation primarily because they are available for all three groups with a uniform level of quality.

\section{Morphometrics}
\label{s:morphometrics}

% rephrase and condense is to derive morphometric parameters:

One observational approach to characterize galaxy appearances is to derive morphometric parameters\footnote{Sometimes called ``non-parametric" as these do not assume a Gaussian distribution of pixel values.}: unitless parameters that do not depend on a preconceived idea about the shape of the profile and are invariant with distance. These morphometric parameters \citep{morphometrics} 
can then be used to classify galaxies along the Hubble Tuning fork or to identify mergers in a population of galaxies \citep{Pearson19,Holwerda23}. 

The morphometric parameters considered here are Concentration, Asymmetry and Smoothness (CAS) from \cite{CAS}, $M_{20}$ and Gini from \cite{Lotz04}, and the Multimode-Intensity-Deviation (MID) parameters from \cite{Peth16}. We use the {\sc statmorph} package described in \cite{Rodriguez-Gomez19} to compute the morphometrics. 

We utilize a Gaussian smoothing kernel with a 1 pixel FWHM (6") for the \HI{} implementations of {\sc statmorph}. This choice is not critical for most computed morphometrics except for the S\'{e}rsic profile fit in {\sc statmorph} and the Intensity and Smoothness morphometric parameters (see section 4.1 and 4.3). \HI{} profiles are typically not well described with such a S\'{e}rsic profile \citep[cf][]{Leroy08,Bigiel11,Wang14,Swaters02,Reynolds23}. We did not anticipate the use of Smoothness or Intensity because of this additional tuning parameter (see section 5.2).
The central position of the galaxy ($x_c$, $y_c$) is re-computed by {\sc statmorph} and the segmentation map is the SoFiA 3D mask \citep{Westmeier21} with the frequency axis collapsed.

\subsection{Concentration-Asymmetry-Smoothness (CAS) Morphometrics}
\label{s:cas}

CAS refers to the commonly used Concentration-Asymmetry-Smoothness space \citep{CAS} for stellar morphological analysis of distant galaxies. Concentration of the light, symmetry around the centre and smoothness as an indication of substructure.

Concentration is defined by \cite{Bershady00} as:
\begin{equation}
C = 5 ~ \log (r_{80} /  r_{20})
\label{eq:c}
\end{equation}
\noindent with $r_{f}$ as the radius containing percentage $f$ of the light of the galaxy (see definitions of $r_f$ in \cite{SE,seman}).
In the optical regime (i.e. stellar component), typical values for the concentration index are $C=2-3$ for discs, $C>3.5$ for massive ellipticals, while peculiars span the entire range \citep{CAS}.

The asymmetry is defined as the level of {\em point}-, (or rotational-) symmetry around the centre of the galaxy \citep{Abraham94,CAS}:
\begin{equation}
A = {\Sigma_{i,j} | I(i,j) - I_{180}(i,j) |  \over \Sigma_{i,j} | I(i,j) |  } - A_{bgr},
\label{eq:a}
\end{equation}
\noindent where $I(i,j)$ is the value of the pixel at the position $[i,j]$ in the image, and $I_{180}(i,j)$ is the pixel at position $[i,j]$ in the galaxy's image, after it was rotated $180^\circ$ around the centre of the galaxy. $A_{bgr}$ is an estimate of the contribution of the background to this value.  This is the definition, without the background contribution, used in \cite{Holwerda12c} for \HI{} as line emission does not have a clear background contribution to asymmetry. Because we use the postage stamps extracted by SoFiA for the calculation, we use the definition of asymmetry without the background computation. 

In the {\sc statmorph} implementation, the asymmetry is calculated in the inner 1.5 Petrosian\footnote{The Petrosian radius is one of several definitions to automatically assign a size and aperture to inherently fuzzy galaxies. For a comprehensive treatment on them, see \cite{Graham05a,Graham05b}. A different size measure of R1 \citep[1 $M_\odot / kpc^2$ similar to those proposed by ][]{Trujillo20,Chamba22} may make more sense for \HI{} } radii (typical size of the stellar disk), the background asymmetry is subtracted and A is minimized by moving the center of rotation. Note that the maximum value for the asymmetry is 2 (all pixels off-center) and can be negative if the background asymmetry value is large. We note that we do not subtract a background when using the moment-0 \HI{} maps as these are extracted from the field using a 3D source mask. A background subtraction makes more sense with continuum emission where a substantial contribution to morphometrics can be expected (i.e. optical or ultraviolet emission) as opposed to line emission maps as is the case here. Moreover, the subtraction has already happened in the radio continuum subtraction that was applied to the data-cube prior to \HI{} line extraction. In our case, background subtraction is a separate step in the data reduction process. To obtain a background asymmetry contribution, one would have to combine continuum subtraction, source extraction, and asymmetry computation. \cite{Reynolds20} compute this background component using an empty section of the \HI{} cube with the same shape as the mask. This was more useful for their comparison between different \HI{} surveys. Here, the background contribution would be dominated by the \HI{} mask shape but {\sc statmorph} expects to compute it based on a sky background just outside the aperture.

Asymmetry in \HI{} maps or profiles has shown a lot of promise in recent studies to identify perturbed or disrupted disk galaxies \citep[e.g.][]{Reynolds20,Glowacki22b,Watts23,Holwerda23}.

Inspired by the ``unsharp masking" technique \citep{Malin78b}, Smoothness is defined by \cite{Takamiya99} and \cite{CAS} as:
\begin{equation}
S = {\Sigma_{i,j} | I(i,j) - I_{S}(i,j) | \over \Sigma_{i,j} | I(i,j) | }
\label{eq:s}
\end{equation}
\noindent where $I_{S}(i,j)$ is the same pixel in a smoothed image. What type of smoothing is used has changed over the years. Often a fixed Gaussian smoothing kernel is chosen for simplicity.

The fact that this Smoothness has another input parameter in the form of the size of the smoothing kernel, makes it highly ``tunable'', meaning one gets out exactly what the parameter was optimized for. It is very difficult to reliably compare between catalogs and especially samples over different distances. 
The kernel employed here is a Gaussian with a width of 2.5 pixels in the moment0 map. This is less than the beam size of the instrument in question. Thanks to the lower dynamical range in \HI{} maps, one does not expect the high-contrast areas such as HII regions in star-forming galaxies. The smoothing kernel choice is therefore a conservative choice (low amount of smoothing) for the Smoothness parameter. The Smoothness parameter is expected to be less useful in \HI{} than in optical or ultraviolet imaging. 

\subsection{Gini and $M_{20}$}
\label{ss:gm20}

\cite{Abraham03} and \cite{Lotz04} introduce the Gini parameter to quantify the distribution of flux over the pixels in an image.
They use the following definition:
\begin{equation}
G = {1\over \bar{I} n (n-1)} \Sigma_i (2i - n - 1) I_i ,
\label{eq:g}
\end{equation}
\noindent $I_i$ is the value of pixel $i$ in an ordered list of the pixels, $n$ is the number of pixels in the image, and $\bar{I}$ is the mean pixel value in the image. 

The Gini parameter is an indication of equality in a distribution \citep[initially an economic indicator][]{Gini12,Yitzhaki91}, with G=0 the perfect equality (all pixels have the same fraction of the flux) and G=1 perfect inequality (all the intensity is in a single pixel). Its behaviour is therefore in between that of a structural measure and concentration. Gini appears quite sturdy as it does not require the center of the object to be computed. It remains relatively unchanged, even when the object is lensed \citep{Florian16} and it is popular for this reason. However, it depends strongly on the image's signal-to-noise \citep{Lisker08}; noise forces the inclusion of a lot of low-signal pixels, throwing off the entire distribution. This issue is not noisy data but how it typically affects image segmentation. In essence, noise can add pixels with no fraction of the flux in them, artificially increasing the Gini value. However, with a less concentrated radial profile and choices of segmentation already made by SoFiA, Gini is a good fit for \HI{} maps. 

\cite{Lotz04} also introduced a way to parameterize the extent of the light in a galaxy image. They define the spatial second order moment as the product of the intensity with the square of the projected distance to the centre of the galaxy. This gives more weight to emission further out in the disk. It is sensitive to substructures such as spiral arms and star-forming regions but insensitive to whether these are distributed symmetrically or not.
The second order moment of a pixel $i$ is defined as:
\begin{equation}
M_i = I_i \times [(x-x_c)^2 + (y-y_c)^2 ],
\label{eq:Mi}
\end{equation}
where $[x, y]$ is the position of a pixel with intensity value $I_i$ in the image and $[x_c, y_c]$ is the central pixel position of the galaxy in the image. 

The total second order moment of the image is given by:
\begin{equation}
M_{tot} = \Sigma_i M_i = \Sigma I_i [(x_i - x_c)^2 + (y_i - y_c)^2].
\label{eq:mtot}
\end{equation}

\cite{Lotz04} use the relative contribution of the brightest 20\% of the pixels to the second order moment as a measure of disturbance of a galaxy
after sorting the list of pixels by intensity ($I_i$):
\begin{equation}
M_{20} = \ log \left( {\Sigma_i M_i  \over  M_{tot}}\right), ~ {\rm for} ~ \Sigma_i I_i < 0.2 I_{tot}. \\
\label{eq:m20}
\end{equation}
The $M_{20}$ parameter is sensitive to bright regions in the outskirts of disks and higher values can be expected in galaxy images (in the optical and UV) with star-forming outer regions as well as those images of strongly interacting disks. Due to a lack of high contrast clumps at higher radii, the $M_{20}$ parameter is not expected to show as much of a range in \HI{} compared to star-formation dominated wavelengths where it was first employed. 

\subsection{Multimode–Intensity–
Deviation (MID) morphometrics}
\label{ss:mid}

The MID morphometrics \citep{Freeman13,Peth16} were introduced as an alternative to the Gini–M20 and CAS morphometrics to be more sensitive to recent mergers. However, these new morphometrics have not been tested as extensively as the Gini–M20 and CAS statistics, especially using hydrodynamic simulations \citep{Lotz08a, lotz10a,Lotz11b, Bignone17}, see also the discussion in the implementation in {\sc statmorph} \citep{Rodriguez-Gomez19}. In the case of \HI{} data for the Hydra cluster, these parameters did not contribute new information \citep{Holwerda23}.

The multimode statistic (M) measures the ratio between the areas of the two most ``prominent'' clumps within a galaxy. The implicit assumption is that the galaxy is well resolved and has at least two well-defined clumps. 
Its calculation mostly consists in finding such substructures. First, all pixels within the MID segmentation map are sorted by brightness. Then, for a given quintile $q$ (between 0 and 1), the set of all pixels with flux values above the $q$th quintile will generally consist of $n$ groups of contiguous pixels, which are sorted by area (largest first). 
Finally, $M$ is defined as the quintile q that maximizes the area ratio between the two largest groups \citep{Peth16}:
\begin{equation}
    \label{eq:M}
    M = max\left({A_{q,2} \over A_{q,1}}\right)
\end{equation}{}
where $A_{q,1}$ is the largest quintile and $A_{q,2}$ is the second-to-largest quintile area. 

% \subsection{Intensity (I)}
% \label{ss:intensity}

The intensity statistic ($I$) measures the ratio between the two brightest subregions of a galaxy. To calculate it, the galaxy image is first slightly smoothed using a Gaussian kernel with $\sigma = 1$
pixel. Then, the image is partitioned into pixel groups according to the watershed algorithm: each distinct subregion consists of all the pixels such that their maximum gradient paths lead to the same local maximum. Once the pixel groups are defined, their summed intensities are sorted into descending order: I1, I2, etc. The intensity statistic is then defined as \cite{Freeman13}:
\begin{equation}
    \label{eq:I}
    I = {I_2 \over I_1}
\end{equation}{}
The same issue that can be raised for M can be raised here. There is a built-in assumption of resolved structure and that this structure has not fractured the segmentation map into separate catalog entries.

The deviation statistic ($D$) measures the distance between the image centroid, ($x_c,y_c$), calculated for the pixels identified by the MID segmentation map, and the brightest peak found during the computation of the $I$ statistic, ($x_I$ , $y_I$ ). This distance is normalized by $\sqrt{n_{seg}/\pi}$, where $n_{seg}$ is the number of pixels in the segmentation map, which represents an approximate galaxy ``radius'' \cite{Freeman13}:
\begin{equation}
    D = \sqrt{\pi \over n_{seg}} \sqrt{ (x_c - x_I)^2 + (y_c - y_I)^2}
\end{equation}{}
This is a metric that can be calculated from Source Extractor \citep{Bertin96,seman} output by using the image centroid and the peak location, albeit again that this makes assumptions on the number of substructures in the galaxy image.

\subsection{Patchiness}
\label{ss:patchiness}

A recent addition to the morphometric parameter space is a ``patchiness" parameter \citep{Fetherolf23} defines as:
\begin{equation}
P = -log_{10} \left\{ \Pi^N_i {1 \over \sqrt{2\pi\sigma_i}} exp \left[ - {(X_i-\bar{X_w})^2 \over 2\sigma_i^2} \right] \right\}
\end{equation}
\noindent where $N_i$ is the number of pixels, $X_i$ is the  value of pixel $i$. The Gaussian probability that all the pixels equal the weighted average is lower when the image is ``patchier". Here, $X_w$  is the weighted (or not) average of the distribution of pixels that make up the object and $\sigma_i$ is the pixel uncertainty. 
The benefits are that this measure is sensitive to deviations above and below the average. It is also notable that this parameter, like the Gini parameter, does not depend on the central position, unlike $M_{20}$ or Asymmetry, which relies on a bright subset of pixels and the object's central position. \cite{Fetherolf23} use their parameter for Voronoi tesselations of their objects and not individual pixels. 
However, we implement a pixel-based definition here. The implementation in \cite{Fetherolf23} focused on their reddening maps i.e., the dust distribution. 
Therefore, this seemed a likely \HI{} morphometric. Upon implementation however, it became clear that the values computed from SoFiA maps are often infinite. For completeness, we include the values in our final catalogue, but not use it in the kNN training below.

\subsection{S\'{e}rsic Profile}

The final step by {\sc STATMORPH} is to fit a single S\'{e}rsic profile with the effective radius ($r_{50}$) and index ($n$) to the pixel collection constituting each object. This is not the optimal description of the \HI{} disk which is usually described with a $R_{1 M_\odot}$, the radius where the profile reaches $1 M_\odot/pc^2$ in \HI{} mass. However, this alternate morphometric, very commonly used in optical studies, is available for use here, and we include the S\'{e}rsic index for consideration.

\subsection{STATMORPH}
\label{ss:statmorph}

Calculating catalogues for these WALLABY data is straightforward for the cutouts provided by the WALLABY data-release. One can run through all the entries made by the SoFiA source detection and run {\sc statmorph} \citep{Rodriguez-Gomez19}. These catalogs are our starting point for the machine learning approach described in the following sections.

\begin{figure*}
    \centering
    \includegraphics[width=\textwidth]{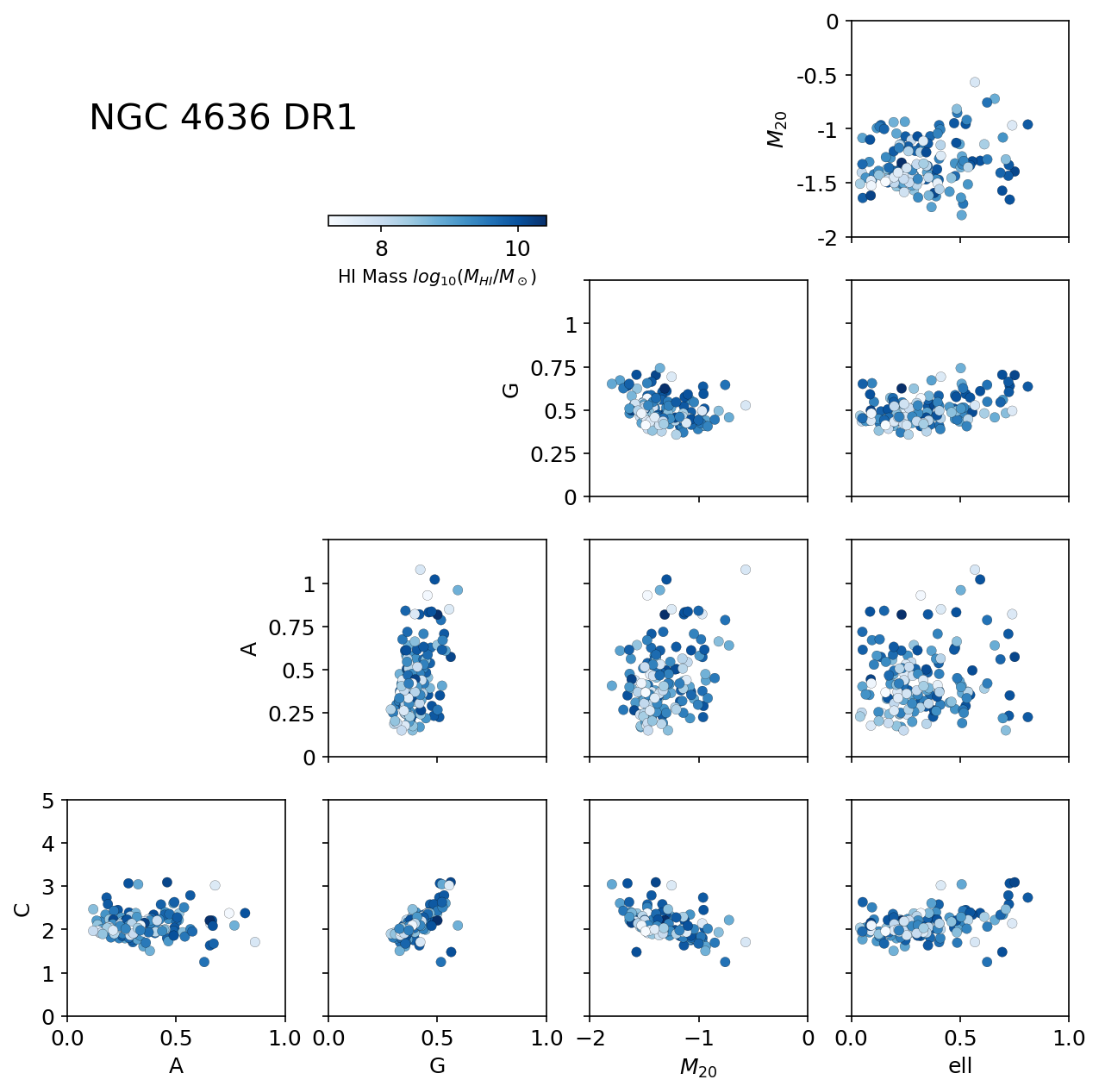}
    \caption{A corner plot of the \HI{} morphometrics of galaxies in the NGC 4636  pointing based on the SoFiA segmentation maps.}
    \label{f:corner:n4636}
\end{figure*}

\begin{figure*}
    \centering
    \includegraphics[width=\textwidth]{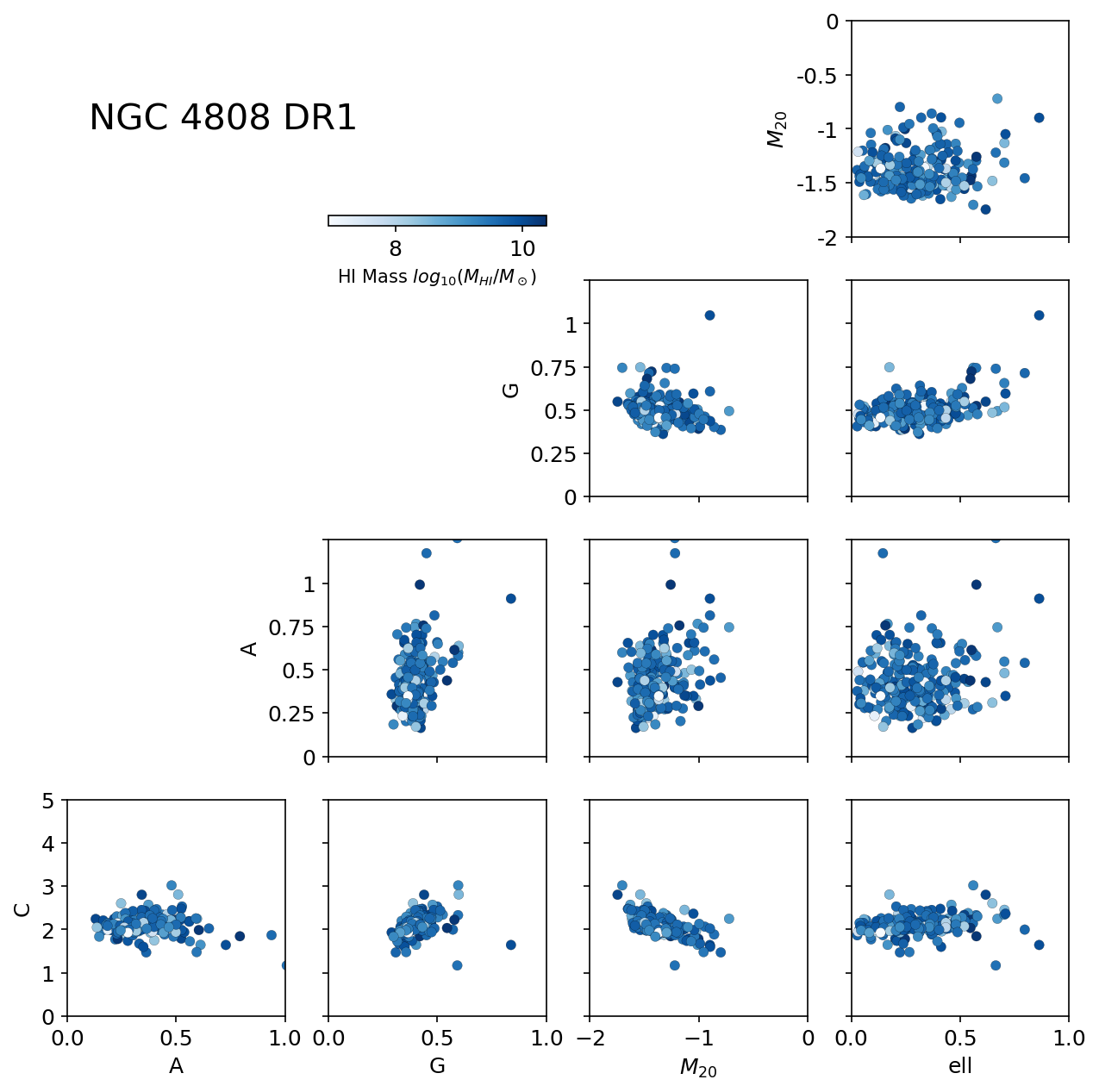}
    \caption{A corner plot of the \HI{} morphometrics of galaxies in the NGC 4808  pointing based on the SoFiA segmentation maps.}
    \label{f:corner:n4808}
\end{figure*}

\begin{figure*}
    \centering
    \includegraphics[width=\textwidth]{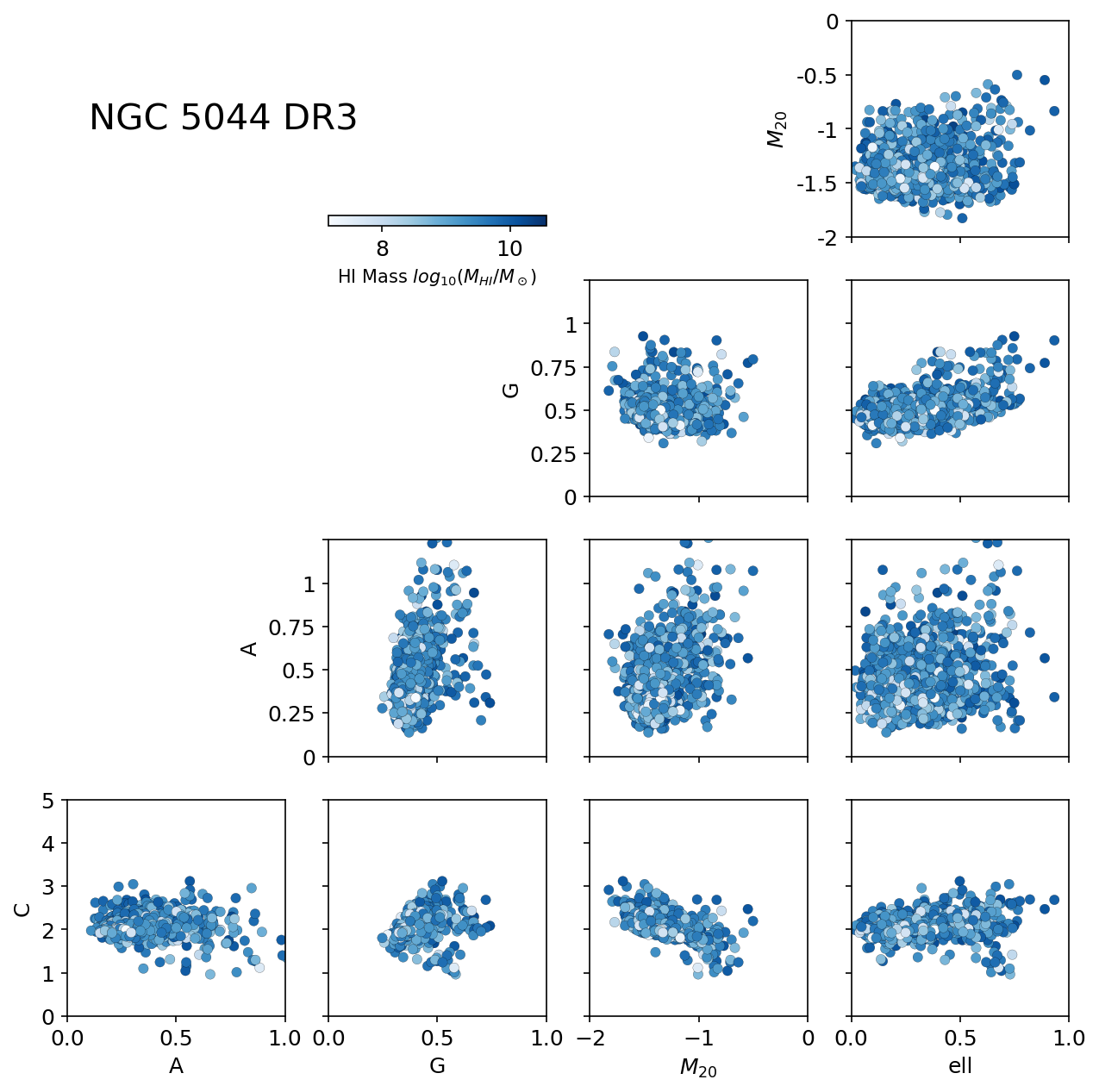}
    \caption{A corner plot of the \HI{} morphometrics of galaxies in the NGC 5044  pointing based on the SoFiA segmentation maps.}
    \label{f:corner:n5044}
\end{figure*}

Part of the parameter space of \HI{} morphometrics are presented in Figures \ref{f:corner:n4636} through \ref{f:corner:n5044}, colour-coded by the inferred \HI{} mass from the SoFiA catalogue. The \HI{} masses are derived after application of the \HI{} flux correction as described in \cite{Westmeier22}. This flux correction is a critical step to match the \HI{} size-mass relation. Concentration, Asymmetry, Gini and $M_{20}$ are the most commonly used parameters.

These are full morphometric catalogs for each field, that is, all galaxies at all redshifts. This approach gives us a sense of the range of values expected. For the subsequent analysis, we apply a cut of $D < 60$ Mpc to select galaxies at mostly similar distances \citep[similar to the samples in ][]{Reynolds20, Holwerda11b}. This distance cutoff ensures the larger features in the \HI{} disks are included in the morphometric calculation; WALLABY's spatial resolution of 30'' $\simeq 10$kpc at this distance. We intentionally do not select known group members because eventually we hope to apply this technique on WALLABY blindly, without prior knowledge of group membership.

Because we do not have full intuition which morphometrics are the optimal feature space to train a machine learning algorithm on --even after the initial work in \cite{Holwerda23}-- we start with the full morphometric space provided by {\sc statmorph}. We do know that Smoothness and Intensity are likely too dependent on the smoothing kernel to be of use in this lower spatial resolution data (see section 4).
This in a way is limiting since there could be other morphometrics much better suited for the identification of perturbed \HI{} disks. It could be possible to define entirely new ones, perhaps including kinematic information as well \citep[cf][]{Deg23}. For now, we adopt the morphometric space provided by {\sc statmorph} with our addition of Patchiness.

Figures \ref{f:corner:n4636} through \ref{f:corner:n5044} show corner plots of the most commonly used morphometrics \citep[modeled after the corner plot in][]{Scarlata07a}. There are some correlations between Concentration and Gini or Concentration and $M_{20}$ evident, something noted by \cite{Conselice08} and \cite{Lotz08}. This morphometric space is not an orthogonal one, especially not with lower resolution data. An orthogonal space would be the easiest to train a machine learning algorithm on and engineer the feature space.
Thanks to a large body of work applying these morphometrics to data from ultraviolet through radio wavelengths, the morphometric space is a familiar one to astronomy.

\section{Machine Learning}
\label{s:ML}

Our approach to these data-sets is to use the objects in the two Virgo fields (NGC 4808 and NGC 4636) as the training set. We have a series of labels for this set from \cite{Lin23d} which can be converted to a simplified flag. Trained on the training sample, we can then exploit first how well classification works (train and test) and then deploy the classifier on the other galaxies in and near these three groups. 

\begin{figure*}
    \centering
    \includegraphics[width=\textwidth]{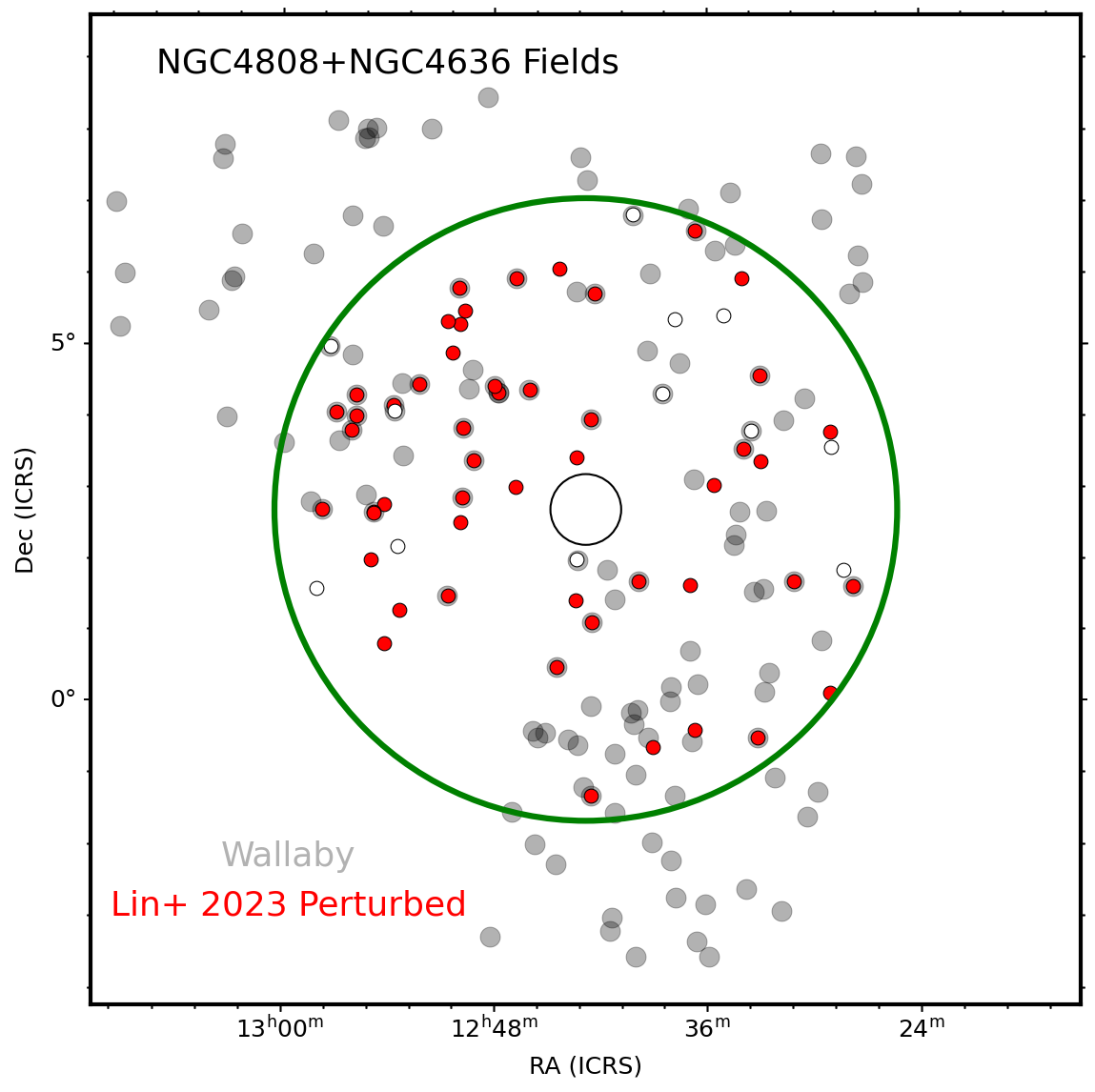}
    \caption{The WALLABY detections for both the NGC 4808 and NGC 4636 fields within ($60$ Mpc) in gray. Superimposed is the catalogue from \cite{Lin23d} with the perturbed (red circles) and unperturbed (white circles). Not every source in \cite{Lin23d} has a counterpart in the two WALLABY catalogues but a sufficient number is available for training. Because the \cite{Lin23d} catalogue is based on different data, we select all the sources within the green circle to be used as the WALLABY training sample with those without a \cite{Lin23d} classification deemed ``unperturbed''.}
    \label{f:map:training}
\end{figure*}

\subsection{Training Sample}
\label{s:training}

To construct a training sample, we require WALLABY \HI{} morphometrics and a label. For the labeling, we use the sample from \cite{Lin23d} who classified galaxies in this field using FAST and WALLABY information. We crosscorrelated the catalogue of \cite{Lin23d}
with both the NGC 4636 and NGC 4808 fields using an arcsecond. We found an overlap with the 63 sources from \cite{Lin23d} with the WALLABY catalogues of 21 and 15 sources in the NGC 4636 and NGC 4808 fields, respectively. We impose our distance limit of 60 Mpc, arriving at a training sample with \cite{Lin23d} labels of 29 sources and 57 WALLABY sources without a label but within that distance and the area of the catalogue (the green circle in Figure \ref{f:map:training}). These unlabeled WALLABY sources are considered to be ``non-perturbed''. Combined, these form our training sample

The final training sample is 29 WALLABY sources with some sort of perturbance and 57 WALLABY sources without the perturbed label. This is a reasonably size and balanced training/test sample which can be complemented using {\sc smote}\footnote{A common technique to balance a training set by re-sampling the under-represented label.} \citep[Synthetic Minority Oversampling Technique,][]{SMOTE} to fully balance the training sample.
The galaxies outside the green circle in Figure \ref{f:map:training} as well as all the objects in the NGC 5044 catalogue are our ``deployment'' sample: the sources the trained algorithm will be deployed on for independent classification. 

\begin{figure*}
    \centering
    \includegraphics[width=0.65\textwidth]{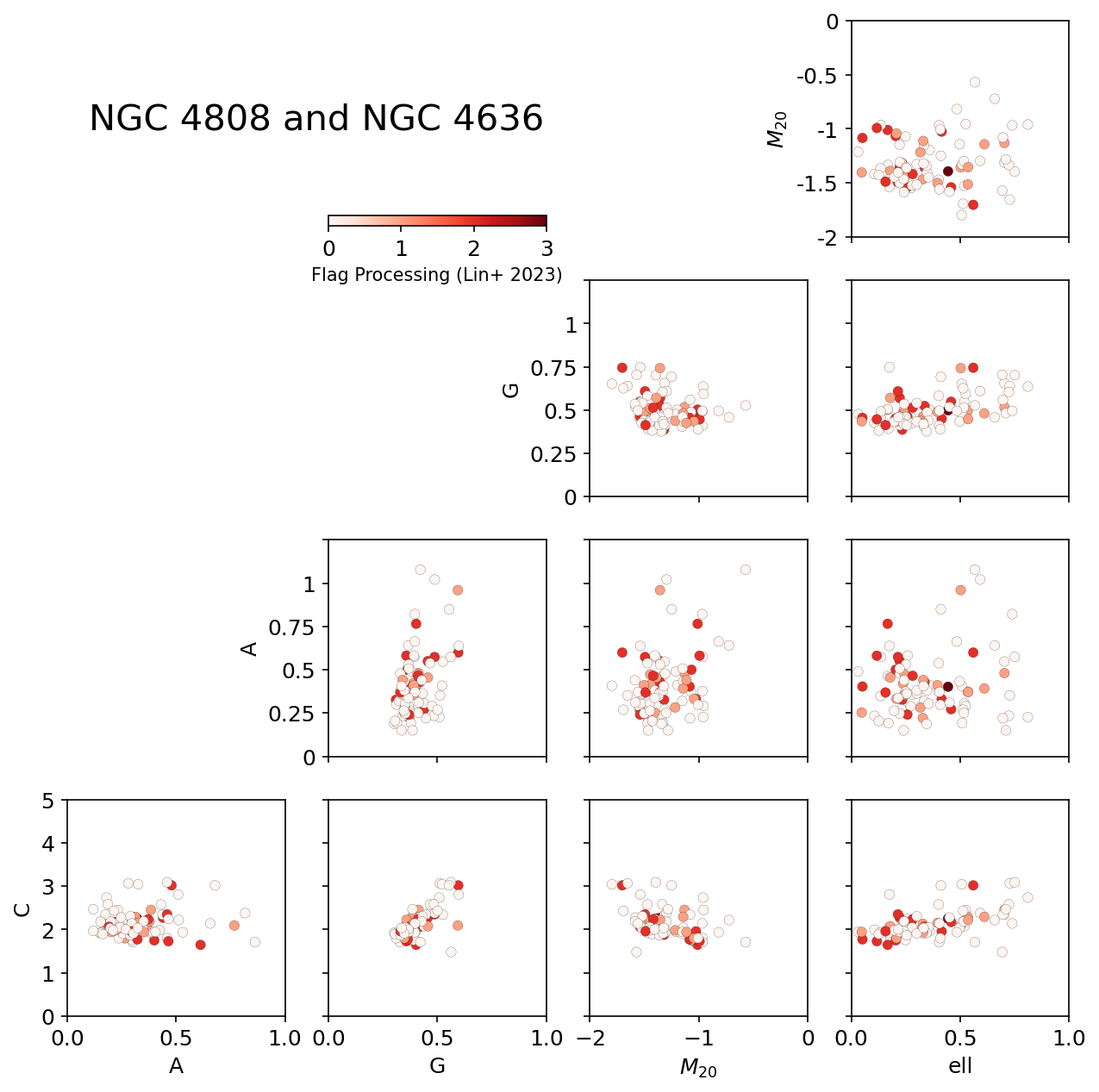}
    \includegraphics[width=0.65\textwidth]{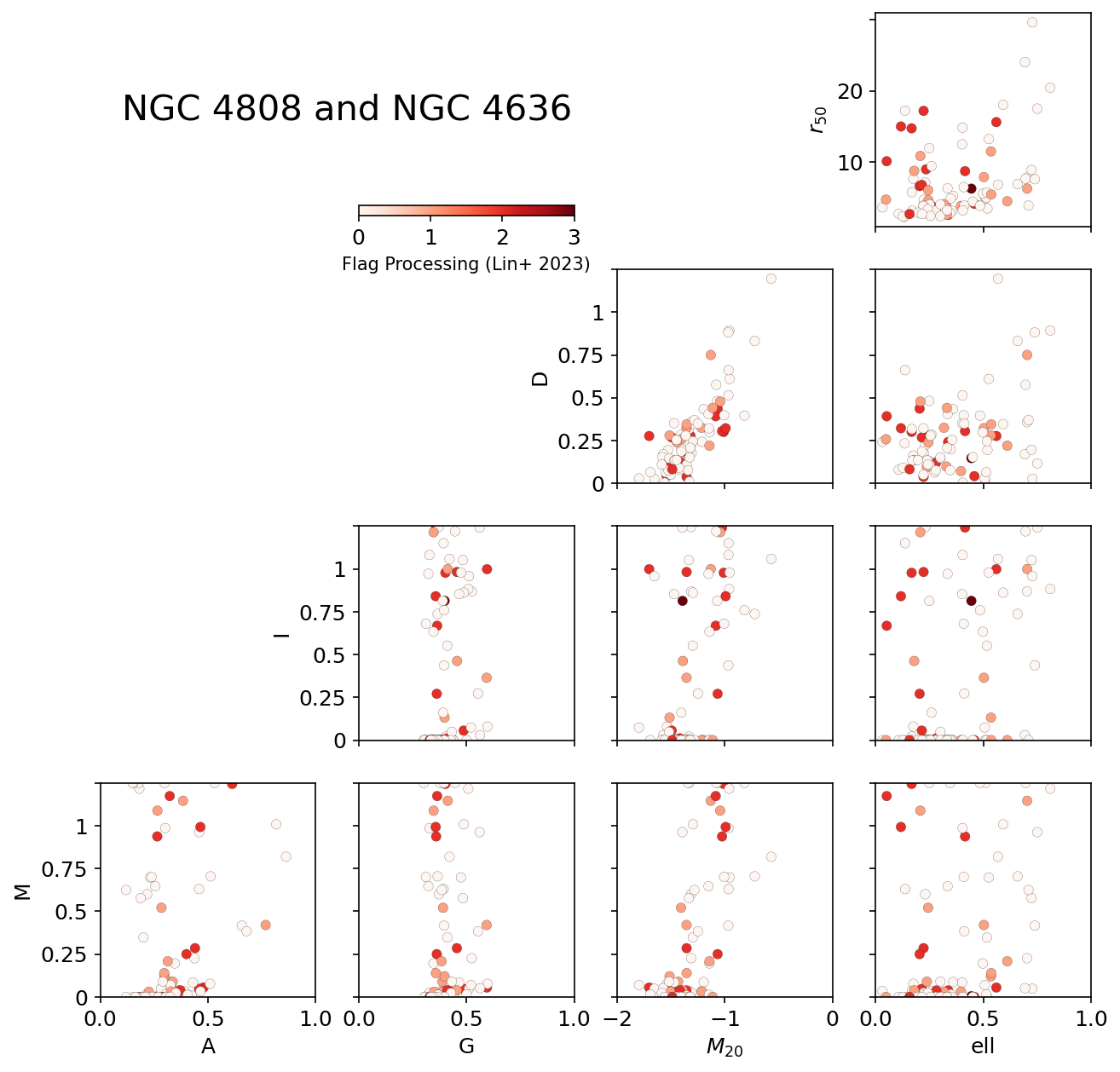}
    \caption{The processing flag for WALLABY objects according to \cite{Lin23d}: ram-pressure (1), tidal interaction (2) or merger (3). We train kNN to distinguish between an undisturbed (0) and processing (1) label which includes all three here (1-3). }
    \label{f:corner:n4808:fproc}
\end{figure*}

Figure \ref{f:corner:n4808:fproc} shows the \HI{} morphometric feature space with the label from \cite{Lin23d} for the galaxies that are undergoing ram-pressure stripping (flag=1), a tidal interaction (flag=2), or a gravitational merger (flag=3). 
The perturbed sample is spread throughout the full \HI{} morphometric space, preempting any possibility to simple cuts in parameter space to separate the two labels. As noted above, the morphometric feature space is degenerate. 

With a limited size training sample, a feature set that is degenerate and no good preset hyperparameter for the ML algorithm (the number of neighbors in this case), we will explore the feature engineering and hyperparameter settings, first separately and then combined. 

For the metrics on performance we will use precision, recall, and F1. Starting with True Positive (TP), True Negative (TN), False Positives (FP), and False Negatives (FN), precision is defined as: $precision = {TP \over TP+FP}$ and recall as: $recall = {TP \over TP+FN}$. F1 is a combination of these: $F1 = 2 \times {Precision \times Recall \over Precision + Recall }$.

\subsection{Feature Engineering}

Because of the size of the training set, we must be extra careful to select a feature space from the \HI{} morphometrics available. Because the undisturbed and perturbed galaxies lie well-mixed in the \HI{} feature space, a k-Nearest Neighbor or a Random Forest (RF) make the most sense to test on this feature space. Each iteration, before we train on this set, we apply {\sc smote} to balance and then the built-in {\sc StandardScaler} in {\sc sk-learn} to whiten (normalize) the data.

First, we examine how many features we will need. Naively, one would use the full \HI{} morphometric space but there is a point of limited return as this is a known degenerate parameter space \citep[see][]{Scarlata07a}. At some point, one would no longer provide new information, just artificially weigh in on features already provided in another format. 
If we use the built-in function {\sc SelectKBest} in {\sc sklearn}, and ask for the 6 highest performing features for the interaction label, we arrive at Concentration, Gini, $M_{20}$, Multimode, Deviation, and S\'{e}rsic index (n). This validates our initial suspicion that Smoothness and Intensity do not hold much additional information in this data and are too dependent on the smoothing kernel size.

\begin{figure}
    \centering
    \includegraphics[width=\textwidth]{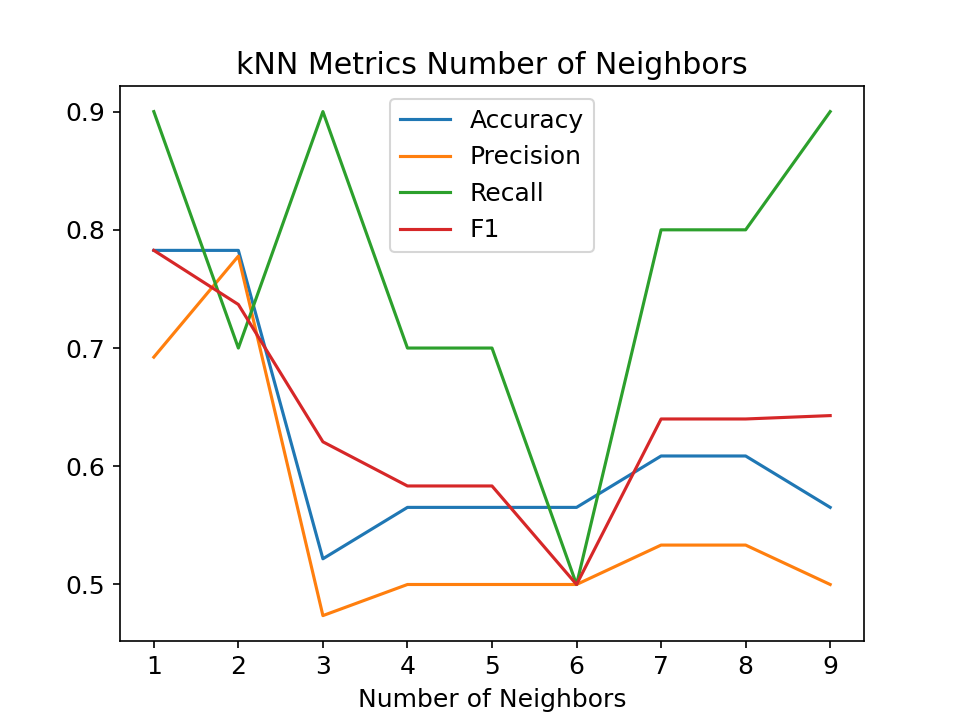}
    \caption{Hyperparameter choice for a set feature space with metrics as a function of the number of neighbours (k). This is the performance for the \textit{full} of morphometric space. }
    \label{f:knn:Nneighbours}
\end{figure}

\subsection{Hyperparameter Optimization}

Figure \ref{f:knn:Nneighbours} shows the number of neighbours used and the different metrics. There is a notable intersection at k=2 and k=6 when using the full parameter space. 
Here, the metrics are very similar, while at k=1,3,4 and 5, the trade-off between recall and precision is overly skewed in favor of recall. This is not clearly reflected in $F1 = 2 \times {Precision \times Recall \over Precision + Recall }$. 
Ideally we would keep the number of neighbors low since we are dealing with a small training set. The high number of neighbors (k=6) would average over a large fraction of the training set every time. A single neighbor suffers from high variance in the classification and affect reliability, essentially over-fitting. 

\begin{figure*}
    \centering
    \includegraphics[width=0.49\textwidth]{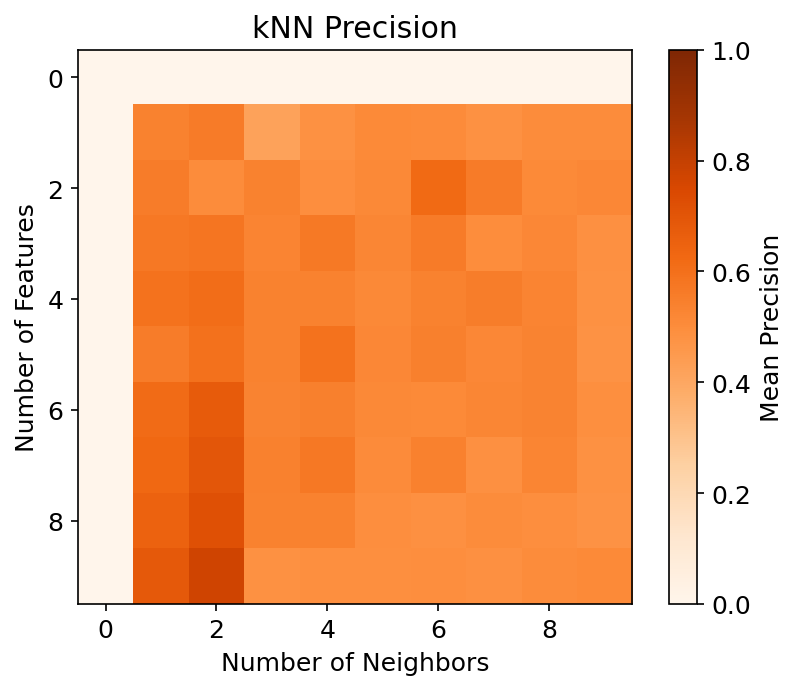}
    \includegraphics[width=0.49\textwidth]{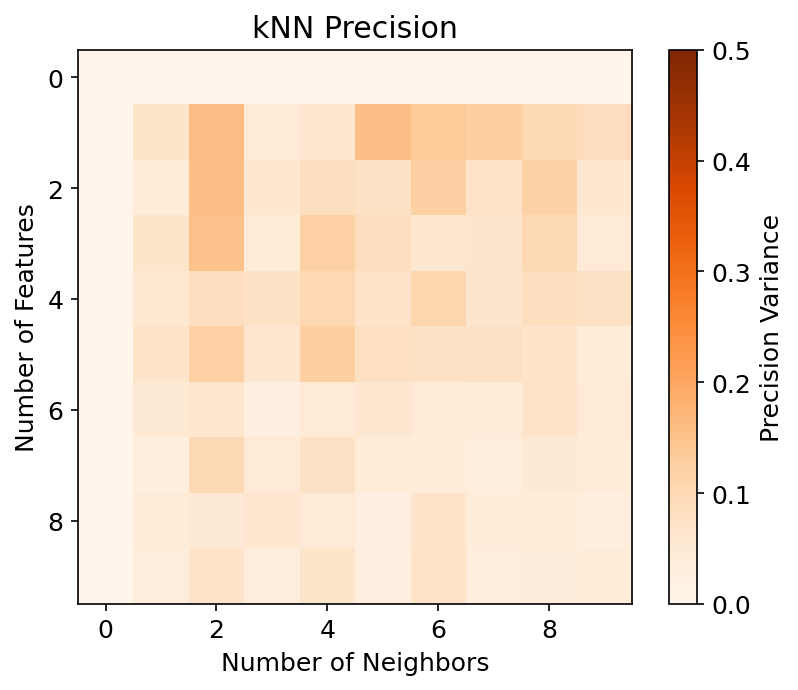}\\
    \includegraphics[width=0.49\textwidth]{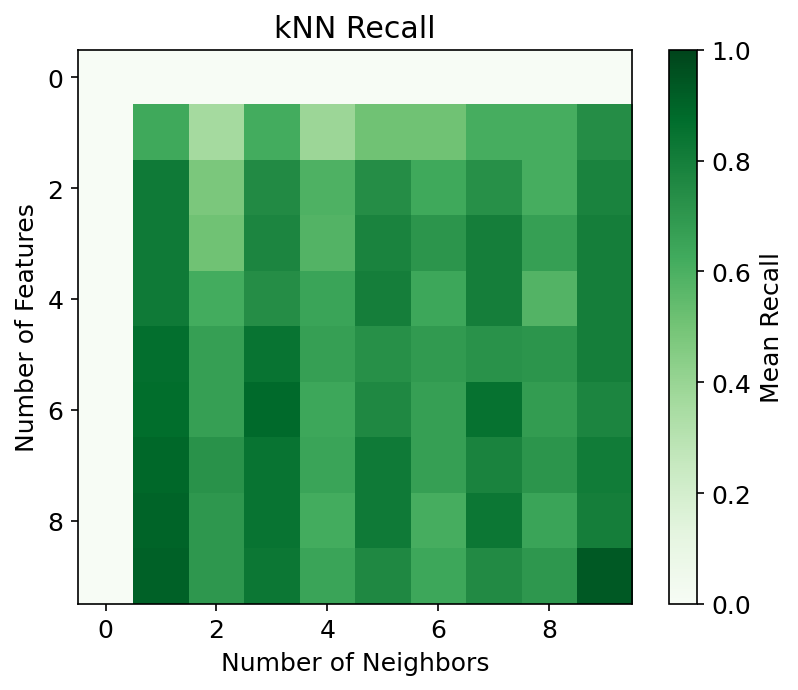}
    \includegraphics[width=0.49\textwidth]{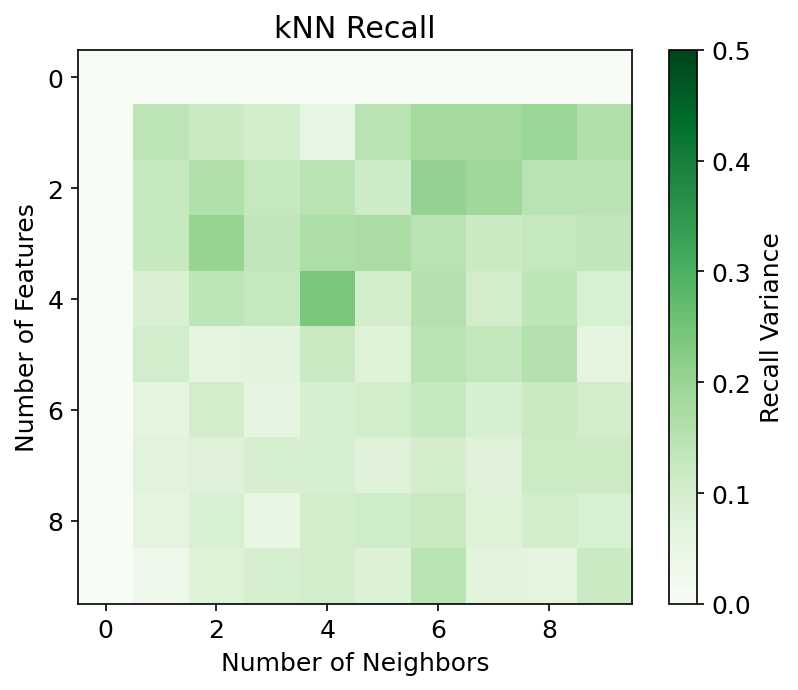}\\
    \includegraphics[width=0.49\textwidth]{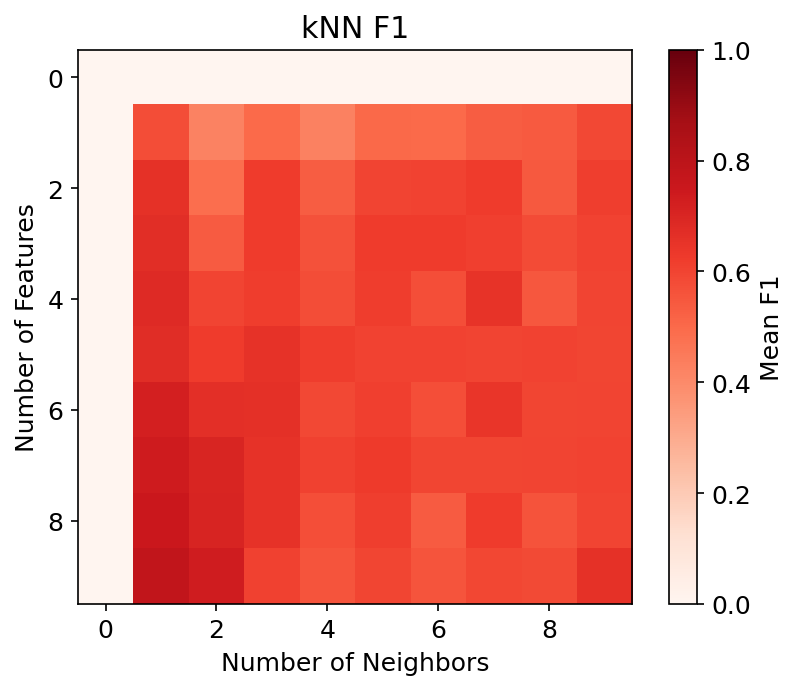}
    \includegraphics[width=0.49\textwidth]{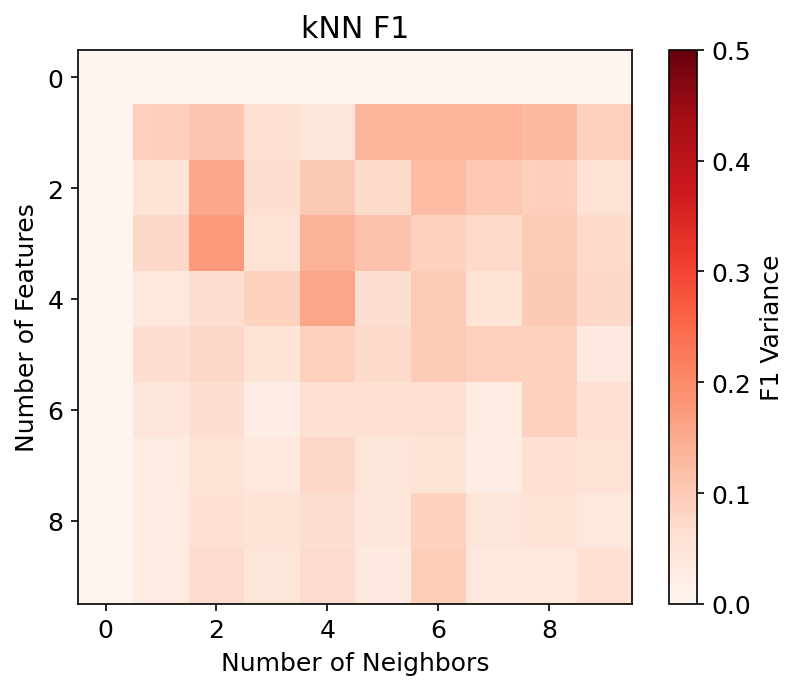}\\
    \caption{The mean (left row) and variance (right row) map of the precision, recall and F1. Mean and variance are determined by drawing a random set of features in the \protect\HI{} morphometric space and running the kNN on it. Variance tends to be high for k=1 or n=2 features.  }
    \label{f:knn:maps}
\end{figure*}
We examine the kNN mean and variance of all the metrics by running multiple iterations with a number of features, randomly selected and a setting for the hyper-parameter (k), the number of neighbours (Figure \ref{f:knn:maps}).
 
Optimization of both the hyperparameter ($k$, number of neighbors) and the feature space, specifically how many features to use, depends on which metric is considered more valuable. Does one want high precision (accurate classifications) or a high recall (reliable classifications) and the F1 metric is meant to reflect a balance between the two. Historically, for merger statistics using morphometrics and other techniques, a high precision was valued since the merger fraction was the aim of the study. However, with more detailed individual galaxy studies, recall may be of higher value for observational follow-up. We therefore aim to strike a balance. 

\begin{figure}
    \centering
    \includegraphics[width=\textwidth]{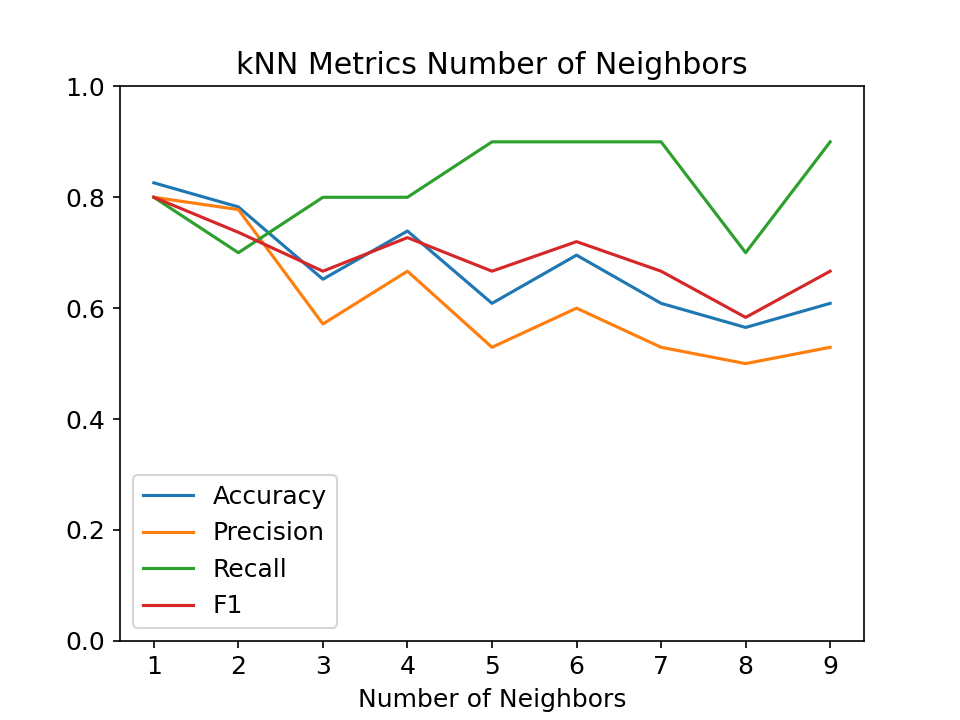}
    \caption{Hyperparameter choice for a set feature space with metrics as a function of the number of neighbours (k). This is for the optimal set of features in Table \ref{t:features}.  }
    \label{f:knn:Nneighbours2}
\end{figure}

To map out the balance between precision and recall here, we map both the mean value and variance as a function of the number of neighbours (k) and the number of features in Figure \ref{f:knn:maps} for each metric. The key here is not that the number of features is increased but that which are used is chosen randomly. So the training set does not automatically start with concentration and moves on from there. The mean value for a combination of neighbours and features tells us how well the kNN algorithm is performing but the variance for that combination (the right side panels in Figure \ref{f:knn:maps}) informs us how reliable that performance is. This is missing in a simpler diagnostic plot such as Figures \ref{f:knn:Nneighbours} or \ref{f:knn:Nneighbours2} which concentrate on just one aspect. 

Given the size of the training sample and feature space (large but not orthogonal), we opt for k=2 neighbors and more than 4 features for optimal performance. This is partly motivated by Figures \ref{f:knn:Nneighbours} and \ref{f:knn:Nneighbours2} but validated when inspecting Figure \ref{f:knn:maps} for low variance in performance. We select those listed in Table \ref{t:features} based on the experience with Hydra, Figure \ref{f:knn:maps}, and which parameters are reported with high F1 scores and close precision/recall scores.

To illustrate the importance of the choice of feature space, Figure \ref{f:knn:Nneighbours2} shows the metrics for the six features selected by {\sc SelectKBest} in {\sc sklearn}. Interestingly, this feature space performs better than the full morphometric space and it is more consistent with metrics. The choice of k=2 is still well motivated as Recall and the other metrics diverge at k=3 and higher. 

\begin{table}[]
    \centering
    \begin{tabular}{c|c}
Feature         & Abbreviation\\
\hline
Concentration   & C \\
Asymmetry       & A \\
Gini            & G \\
M$_{20}$        & M$_{20}$\\
Deviation       & D \\
S\'{e}rsic index & n \\       
\hline
\end{tabular}
    \caption{The features selected for the final iteration of the kNN. }
    \label{t:features}
\end{table}

\begin{figure}
    \centering
    \includegraphics[width=\textwidth]{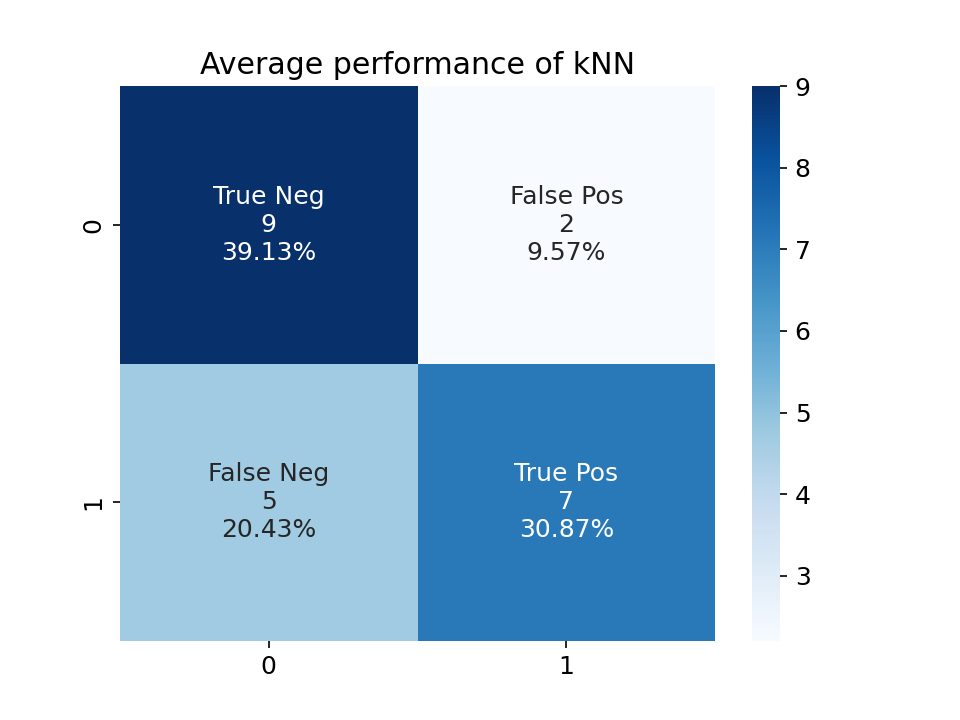}
    \caption{The average confusion matrix for the kNN (k=2, trained on subsamples of 80\%, tested on the remaining 20\% shown here) with the optimized feature space listed in Table \ref{t:features} for all the members of the NGC 4636 and NGC 4636 groups. We repeated the training/test ten times and these are the averages of all ten split-train-test iterations. }
    \label{f:knn:meanconfusion-matrix}
\end{figure}

Similarly, one can argue which six morphometrics are preferred. For example, asymmetry is better understood and more widely adopted than the MID parameters. This could be an argument to include asymmetry instead of the multimode parameter. 
To ascertain the effectiveness of the kNN on this data-set, we evaluate the average of a series training-test runs, where the training/test sample split is 80\%. We do this ten times. This approach is very similar to bootstrapping a simple fit. Figure \ref{f:knn:meanconfusion-matrix} shows the average confusion matrix for the test sample after training on 80\% of total sample. The average metrics of this configuration (the features in table \ref{t:features} and k=2) are listed in Table \ref{t:knn:meanmetrics}. Thanks to the repeat in kNN training/test instances, the metrics also come with a standard deviation around the mean performance. These mean performance metrics are proficient for a simple machine learning algorithm.

\begin{figure}
    \centering
    \includegraphics[width=\textwidth]{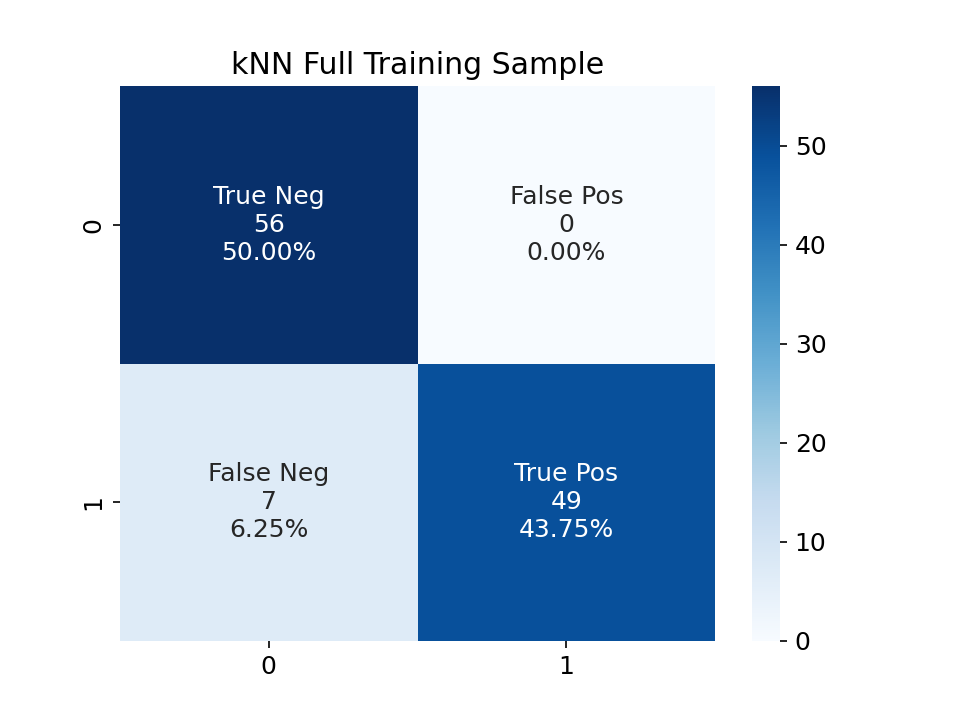}
    \caption{The confusion matrix for the kNN (k=2, trained on a subsample of 80\%) with the optimized feature space listed in Table \ref{t:features} for all the objects in the combined catalogue of the NGC 4808 and NGC 4636 fields (D $<$ 60 Mpc within the green circle in Figure \ref{f:map:training}). }
    \label{f:knn:confusion-matrix}
\end{figure}

However, for application to other data-sets, it can be beneficial to use the entire labeled sample as the training sample. If we do this, the metrics become those in Table \ref{t:knn:metrics} and the confusion matrix in Figure \ref{f:knn:confusion-matrix}. Performance is quite good considering the size of the training sample. We will now employ this kNN (trained on the full labeled sample) on the other catalogue, the one for the NGC 5044 mosaic.

\begin{table}[]
    \centering
    \begin{tabular}{c|c}
    \hline
Accuracy  & 73.04 $\pm$ 6.96  \\
Precision & 79.58 $\pm$ 9.03  \\
Recall    & 66.67 $\pm$ 13.71  \\
F1.       & 71.34 $\pm$ 4.01  \\
        
    \hline
    \end{tabular}
    \caption{The performance metrics of the WALLABY training catalogue (D $<$ 60 Mpc within the green circle in Figure \ref{f:map:training}) split into subsections using the features listed in Table \ref{t:features}. By iterating ten times over this sample and splitting off 20\% for testing, these are the mean and variance of the kNN performance. }
    \label{t:knn:meanmetrics}
\end{table}

\begin{table}[]
    \centering
    \begin{tabular}{c|c}
    \hline
        Accuracy    & 86\% \\
        Precision   & 88\% \\
        Recall      & 70\% \\
        F1          & 78\% \\
    \hline
    \end{tabular}
    \caption{The performance metrics of in the full WALLABY training catalogue (D $<$ 60 Mpc within the green circle in Figure \ref{f:map:training}) using the features listed in Table \ref{t:features}. }
    \label{t:knn:metrics}
\end{table}

\subsection{Biases}

\begin{figure*}
    \includegraphics[height=0.49\textwidth]{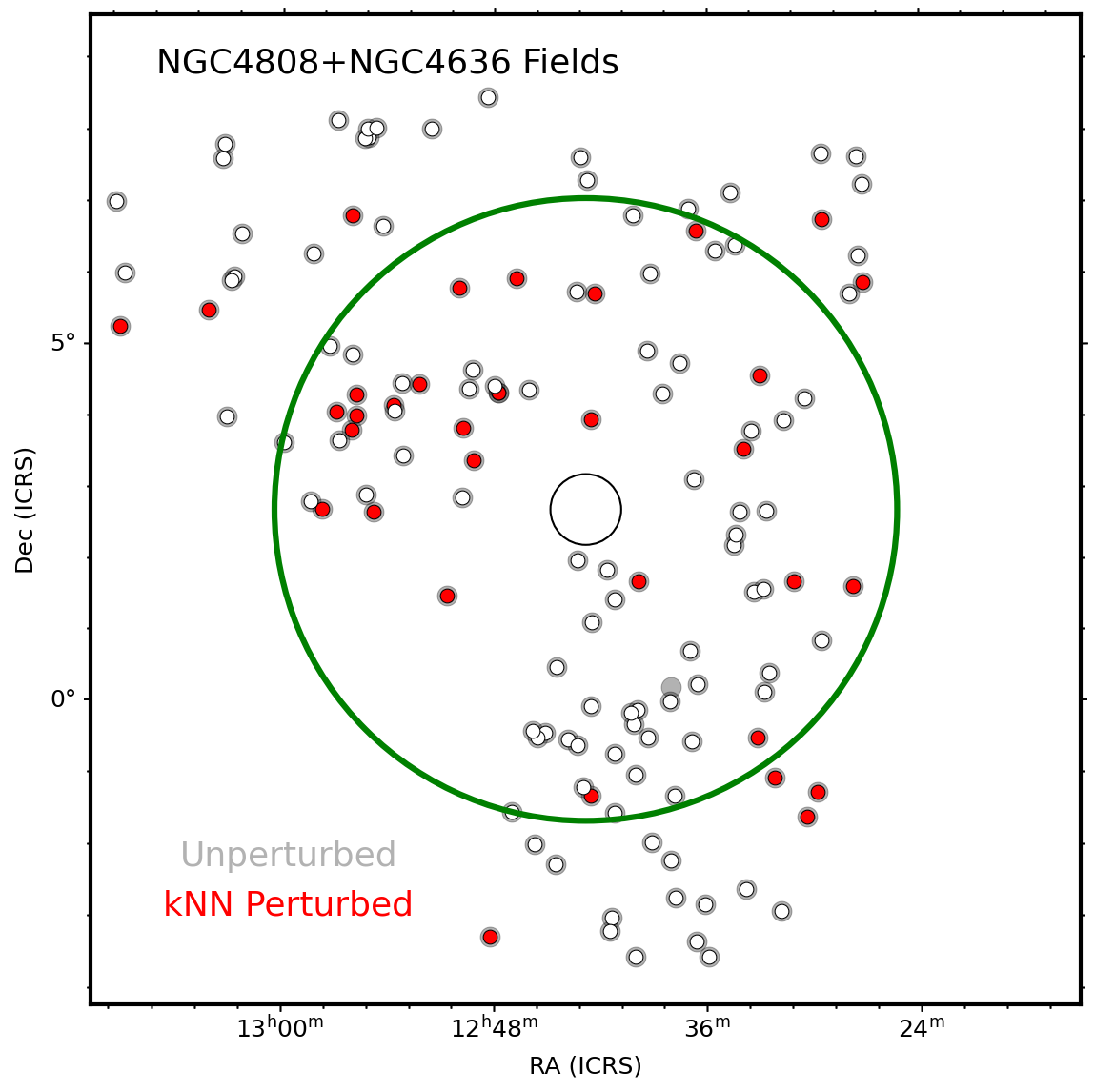}
    \includegraphics[height=0.49\textwidth]{holwerda25a_f12a.png}
        \caption{The kNN labeling in both the training sample (left) and the deployment field, NGC 5044. Compare to the labels in Figure \ref{f:map:training}. }
    \label{f:knnclass:radec}
\end{figure*}

There remains the possibility of biases applying a training sample on a new data-set. The objects in the NGC 5044 mosaic are biased towards greater distances, the signal-to-noise in the different data-cubes varies due to RFI or other factors, etc. The parameterization of morphology through the above morphometrics is meant to be mostly invariant to small changes. Aside from familiarity, this is a prominent reason to convert to morphometrics first before attempting a machine learning algorithm. Our cut in distance to just those galaxies closer than 60 Mpc is also meant to remove biases in the training and application set (e.g. there are many sources in the wide field behind the NGC 5044 group that would skew our results). Figure \ref{f:knnclass:radec} shows the position of the galaxies in each field closer than 60 Mpc with the kNN classification marked.
That said, small differences and thus biases between data-sets may well be present. Based on the distribution of sources in the parameter space, we estimate the issue to be small. However, moving from one sample to another with a (small) fundamental difference is a known issue in machine learning known as ``transfer learning''.

\subsection{Application on NGC 5044}

We cut down the samples to only those galaxies below 60 Mpc for the training sample in the NGC 4808/4636 fields. The NGC 5044 field is a little further away on average but richer and well within this distance limit. The rationale for the distance limit is that it generously includes all labeled galaxies while removing the majority of unresolved background objects (Figure \ref{f:distances}). The resulting sample is 258 galaxies (within $D<60$ Mpc) for the NGC 5044 field. 
In the comparisons in scaling relations, we will compare the training sample scaling relation to this deployment sample of NGC 5044 mosaic. 

\section{Results}
\label{s:results}

\subsection{Fraction of perturbed galaxies}

Table \ref{t:disturbed-frac} lists the fraction of the galaxies below 60 Mpc. in each of the three groups that the kNN trained on the \cite{Lin23d} classifications as perturbed somehow (i.e. ram-pressure stripping, tidally disturbed or merging). We compare these to ``perturbed'' criteria from the literature, similar to \cite{Holwerda11b}.

The fraction of perturbed galaxies in the catalog of \cite{Lin23d} is slightly lower than what we find using kNN. In general, the kNN finds a similar fraction of galaxies perturbed in each field. 
Based on the metrics listed as listed in Table \ref{t:knn:metrics}, one would expect these fractions to be accurate to within a few percentage points. The difference of $\sim$1\% in NGC 4808 is therefore illustrative of what the uncertainty should be. 

\begin{table*}[]
    \centering
    \begin{tabular}{l l l l l}
    & Perturbed percentage & & & \\
criterion                               & NGC 4808  & NGC 4636  & NGC 5044  & Reference \\
Lin+ (2023)                             & 33.33     & \dots     & \dots     & \cite{Lin23d}\\
WALLABY training sample                 & 34.88     & \dots     & \dots     & \\
\hline
kNN                                     & 21.35     & 33.33     & 21.71     & This work.\\

$A>0.38$                                & 35.42     & 26.97     & 27.91     & \cite{CAS}\\
$ G > -0.115 \times M_{20} + 0.384 $    & 6.25      & 4.49      & 4.26      & \cite{Lotz04}\\
$ G < -0.15\times M_{20} + 0.33 $       & 6.25      & 6.74      & 4.65      & \cite{Lotz08}\\
$ A < -0.2 \times M_{20} + 0.25 $       & 95.83     & 87.64     & 88.37     & \cite{Holwerda11b}\\
$ C > -5 \times M_{20} + 3. $           & 0.00      & 0.00      & 0.00      & \cite{Holwerda11b}\\

\hline
    \end{tabular}
    \caption{The fraction of galaxies that were perturbed as reported by \cite{Lin23d} and the kNN trained on NGC 4808+4636 (WALLABY training sample). For comparison, the morphometric criteria for merging or perturbed galaxies from \cite{CAS}, \cite{Lotz04, Lotz08} and \cite{Holwerda11b} are listed as well. }
    \label{t:disturbed-frac}
\end{table*}

Previous uses of morphometrics used a simple criterion to separate perturbed from unperturbed galaxies. \cite{Holwerda11b} reviews these in the context of their use on \HI{} surveys. \HI{} morphology is expected to be perturbed earlier and longer during a gravitational interaction. Table \ref{t:disturbed-frac} lists the fractions of galaxies that meet the various criteria as well. It is notable that a the basic asymmetry criterion ($A>0.35$) identifies a similar percentage as the kNN classifier. Once compared however, that Asymmetry criterion is biased towards false negatives. Since in the past, the goal of morphometric identification of mergers was to identify the merger fractions at different epochs or environments, the kNN approach works certainly well enough on a population. 

\subsection{Galaxy Scaling Relations}

One application of a ML classifier is to rapidly classify galaxies to then examine the galaxy scaling relations for those galaxies undergoing some interaction to those that are not. Here we examine three: the star-forming galaxy main sequence, the \HI{} and stellar mass relation, and the Baryonic Tully-Fisher relation. We also looked at the \HI{} size-mass relation but there is little difference between galaxies marked perturbed and not. The lack of an \HI{} size-mass relation can be attributed to the still relatively low spatial resolution of the WALLABY pilot observations, expected to improve, and relatively simple size measures. 

\subsubsection{Star-forming Main Sequence}

\begin{figure*}
    \centering
    \includegraphics[width=0.49\textwidth]{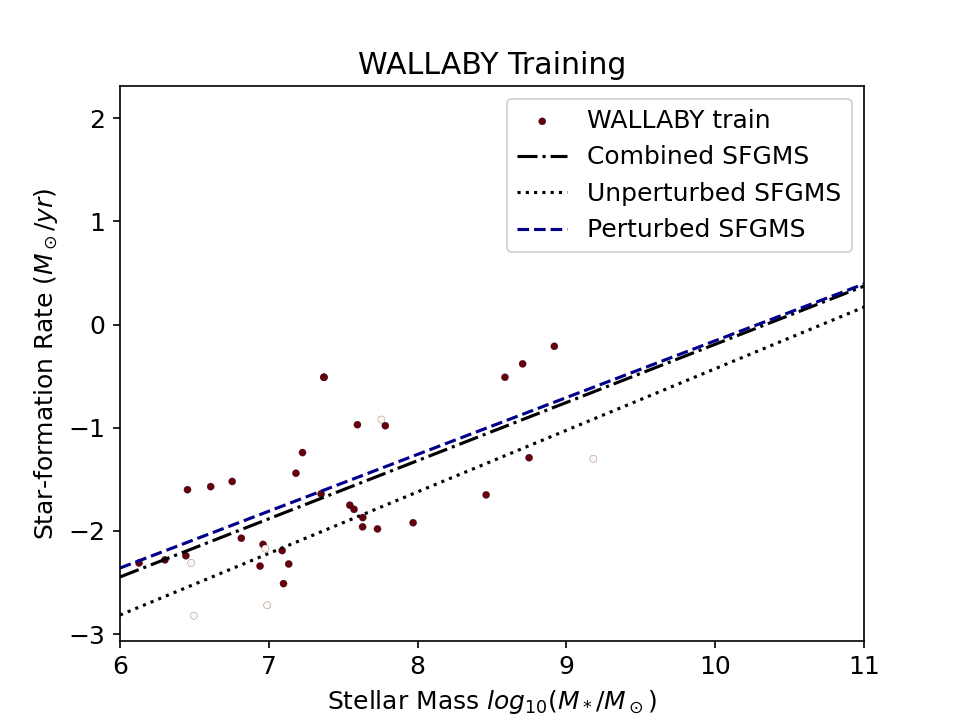}
    \includegraphics[width=0.49\textwidth]{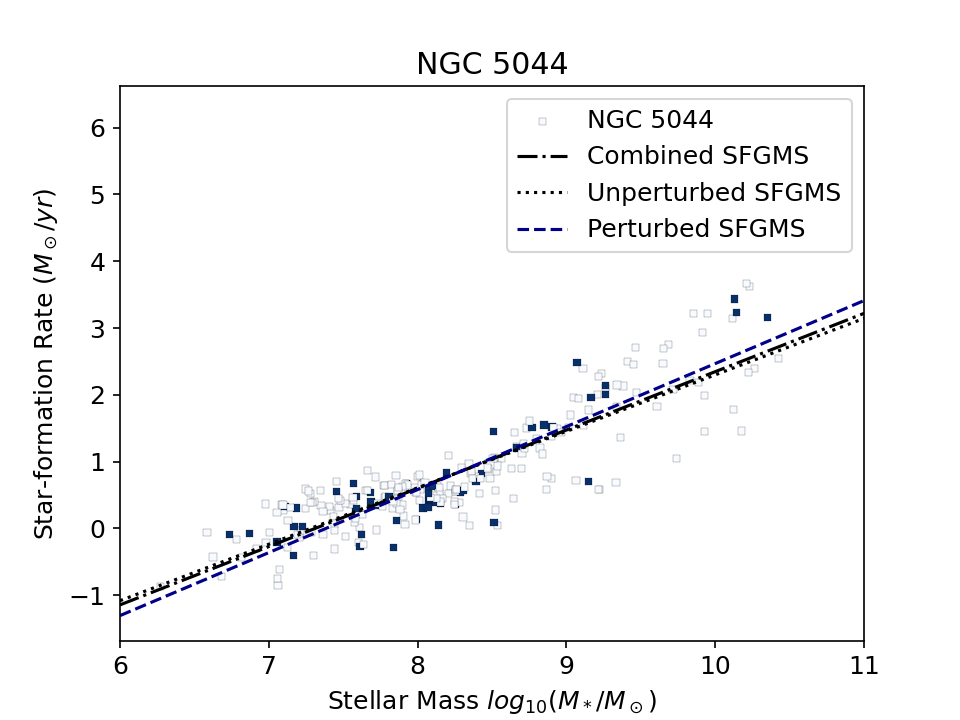}
    \caption{The stellar mass and star-formation relation for the WALLABY training sample (left) and the NGC 5044 deployment sample (right). Qualitatively, the results are similar for the star-forming galaxy main sequence: similar slopes for all three populations, unperturbed, perturbed and all galaxies but there are quantitative differences in the SFGMS slope and interecept between the training and the deployement samples.   }
    \label{f:galscale:SFGMS}
\end{figure*}

The star-forming galaxies main sequence \citep[e.g.,][]{Noeske07} is an important relation between the stellar mass of galaxies and their (relative) growth rate. 

Figure \ref{f:galscale:SFGMS} shows the stellar mass-star-formation relation for the WALLABY training sample and the deployment data in the NGC 5044 mosaic. The kNN-identified perturbed galaxies are mixed in with the main sequence of star-forming galaxies. A linear fit to the stellar mass and star-formation relation for these galaxies, all of whom are on the star-forming main sequence, is essentially the same for perturbed and non-perturbed sets (Table \ref{t:galscale:SFGMS}). We note there is a normalization difference between the training and deployment sample for the SFR estimate from WISE. It is unclear if this is a distance effect, or additional flux in WISE W3 due to Galactic Cirrus. The training and NGC5044 samples show the same slope and intercept within their respective bootstrap errors (Table \ref{t:galscale:SFGMS}).

There is no functional difference in the slope and intercepts between perturbed and unperturbed galaxies.  There is a difference between training and deployment samples but that is to be expected when moving to a sample with a difference mass range.

\begin{table}[]
    \centering
    \begin{tabular}{l l l l l}
Sample          & Training  & & NGC 5044 & \\
                & slope & intercept & slope & intercept \\
\hline
All             & 0.56 $\pm$ 0.11 & -5.83 $\pm$ 0.83 & 0.87 $\pm$ 0.03 & -6.38 $\pm$ 0.28\\
Unperturbed     & 0.60 $\pm$ 0.48 & -6.40 $\pm$ 3.27 & 0.85 $\pm$ 0.04 & -6.15 $\pm$ 0.33  \\
Perturbed       & 0.55 $\pm$ 0.12 & -5.66 $\pm$ 0.88 & 0.94 $\pm$ 0.07 & -6.97 $\pm$ 0.54 \\
\hline
    \end{tabular}
    \caption{The linear fits to the stellar mass and star-formation relation for the training sample and the deployment sample of NGC 5044 for all the galaxies in the sample, the unperturbed and perturbed ones. Qualitatively the fits are similar but the deployment fits have lower slopes and higher intercepts than the training sample. }
    \label{t:galscale:SFGMS}
\end{table}

\subsubsection{Stellar And HI Mass }

\begin{figure*}
    \centering
    \includegraphics[width=0.49\textwidth]{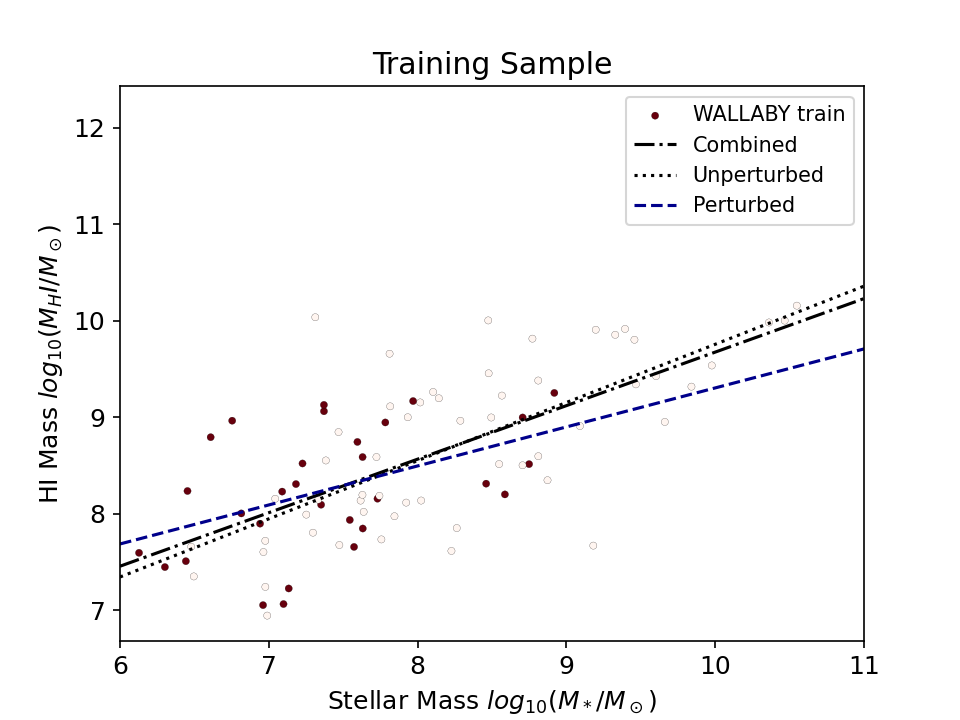}
    \includegraphics[width=0.49\textwidth]{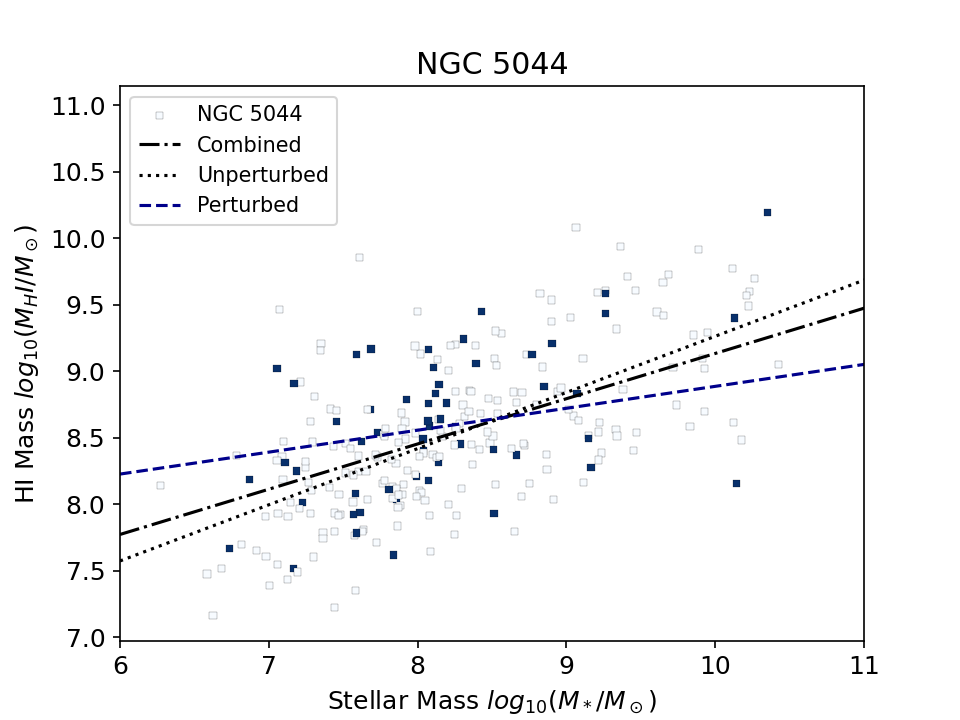}
    \caption{The stellar mass and \HI{} mass relation for the training sample (left) and the deployment sample, the NGC 5044 mosaic. Both the combined and the unperturbed samples show very similar fits and the galaxies indicated as perturbed in the training sample as well as in the NGC 5044 mosaic both show less \HI{} mass for a given stellar mass. }
    \label{f:galscale:MHIMstar}
\end{figure*}

Figure \ref{f:galscale:MHIMstar} shows the stellar mass, as derived from the WISE W1 flux, and the \HI{} mass from the WALLABY catalog for the training sample and the deployment data of the NGC 5044 mosaic. We note that a quantitative comparison with existing relations \citep[e.g.][]{Catinella18} is not done here because stellar mass estimates are based on catalogue photometry on a single filter from WISE. 
Qualitatively, the correlations for this relation are similar for training and deployment samples; unperturbed galaxy relation has a higher slope, the perturbed ones a lower slope than the whole sample fit. Table \ref{t:galscale:MHIMstar} quantifies this with bootstrapped errors. We note here that the training set skews a little lower mass than the deployment sample. 

\begin{table}[]
    \centering
    \begin{tabular}{l l l l l}
Sample          & Training  & & NGC 5044 & \\
                & slope & intercept & slope & intercept \\
\hline
All             & 0.55 $\pm$ 0.05  &  4.15 $\pm$ 0.45 & 0.34 $\pm$ 0.07 & 5.73 $\pm$ 0.57 \\
Unperturbed     & 0.60 $\pm$ 0.07  &  3.76 $\pm$ 0.57 & 0.42 $\pm$ 0.04 & 5.03 $\pm$ 0.32 \\
Perturbed       & 0.40 $\pm$ 0.11  &  5.26 $\pm$ 0.85 & 0.17 $\pm$ 0.16 & 7.19 $\pm$ 1.29 \\
\hline
    \end{tabular}
    \caption{The linear fits to the stellar and \HI{} mass relation for the training sample and the deployment sample of NGC 5044 for all the galaxies in the sample, the unperturbed and perturbed ones. Qualitatively the fits are similar but the deployment fits have lower slopes and higher intercepts than the training sample. }
    \label{t:galscale:MHIMstar}
\end{table}

\subsubsection{Baryonic Tully-Fisher Relation}

\begin{figure*}
    \centering
    \includegraphics[width=0.49\textwidth]{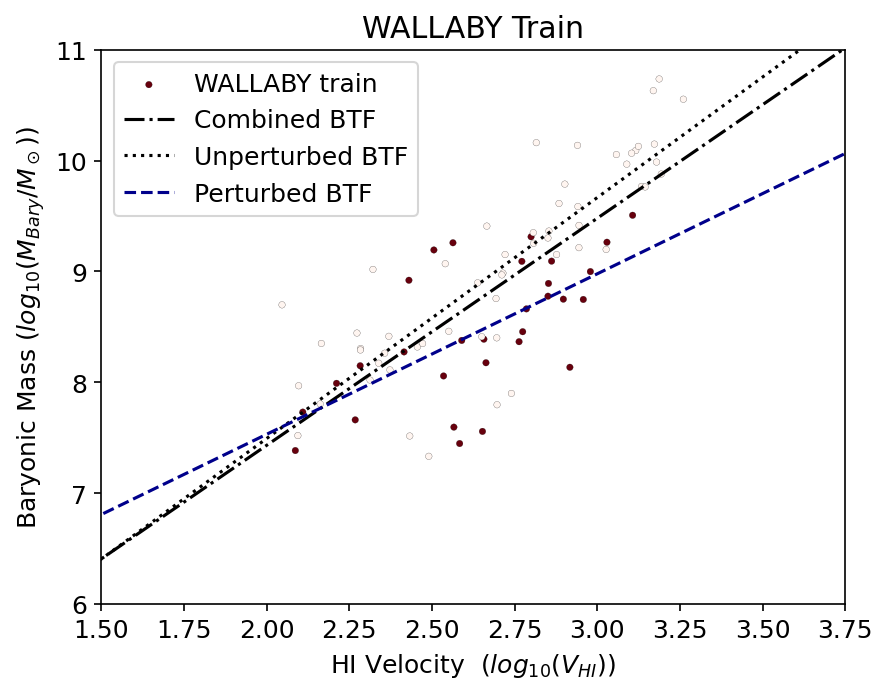}
    \includegraphics[width=0.49\textwidth]{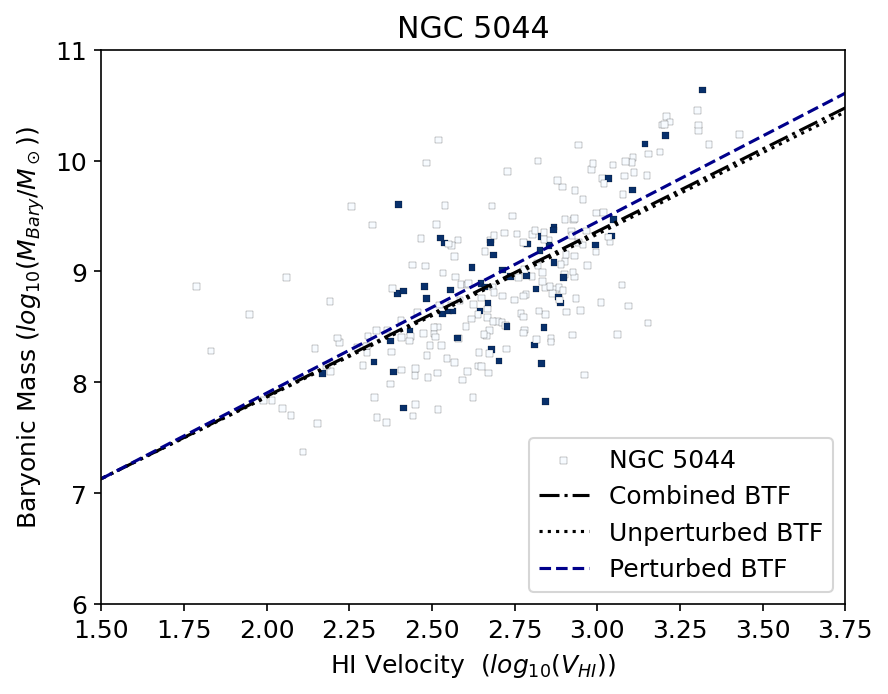}
    \caption{The Baryonic Tully-Fisher relation for the WALLABY training sample (left) and the deployment sample on NGC 5044 (right). The velocity is computed according to equation \ref{eq:VHI} with the unit in m/s.  }
    \label{f:galscale:BTF}
\end{figure*}

Figure \ref{f:galscale:BTF} shows the Baryonic Tully-Fisher relation for all three groups combined. We used the WISE W1 based stellar mass and a factor 1.33 to convert the \HI{} mass into a total gas mass including Helium. The \HI{} velocity is the W50 measurement corrected for inclination from the SoFiA measurements:
\begin{equation}
\label{eq:VHI}
    V_{HI} = { w50 \over \sqrt{1 - \left({b \over a}\right)^2 }}
\end{equation}

The slope and the intercept were fit with a standard linear regression. Because the uncertainty in the Baryonic mass is under-estimated with the formal uncertainties, we estimate the variance in the slope and intercept using a bootstrapping of the fits. The BTF linear fit through all the galaxies and those labeled unperturbed are very similar but the kNN-identified perturbed population shows a flatter BTF relation. The uncertainties reported in Table \ref{t:galscale:BTF} are a standard deviation. The discrepancy is therefore significant. 

The measurements for individual galaxies can have some uncertainties due to model fits. For example, the conversion from WISE W1 flux to a stellar mass and the correction for inclination using the SoFiA measured major and minor axes. Especially in recently perturbed galaxies, this axis ratio may not be indicative of the disk's inherent inclination.

\begin{table}[]
    \centering
    \begin{tabular}{l l l l l}
Sample          & Training  & & NGC 5044 & \\
                & slope & intercept & slope & intercept \\
\hline
All:            & 1.82 $\pm$ 0.14 & -1.32 $\pm$ 0.76     & 1.47 $\pm$ 0.10 & 0.72 $\pm$ 0.56 \\
Unperturbed:    & 1.96 $\pm$ 0.15 & -1.95 $\pm$ 0.84     & 1.45 $\pm$ 0.11 & 0.83 $\pm$ 0.60 \\
Perturbed:      & 1.13 $\pm$ 0.26 & 2.25  $\pm$ 1.46     & 1.54 $\pm$ 0.30 & 0.39 $\pm$ 1.64 \\
\hline
    \end{tabular}
    \caption{The linear fits to the Baryonic Tully-Fisher relation for the training sample and the deployment sample of NGC 5044 for all the galaxies in the sample, the unperturbed and perturbed ones.  }
    \label{t:galscale:BTF}
\end{table}

\section{Discussion}
\label{s:discussion}

\subsection{kNN Performance}

In this paper, we considered only the kNN classification on \HI{} morphometrics, following the experiences in \cite{Holwerda23}. Generally speaking, the perturbed and unperturbed samples are well-mixed in \HI{} morphometric space without a clean separation in this multidimensional space. This is reflected in the low number of neighbours which would optimize classification metrics (Figure \ref{f:knn:Nneighbours}). Once the feature space is optimized, kNN behaviour is much more reasonable but still performs best with two neighbours (see Figure \ref{f:knn:Nneighbours2}).

Overall the kNN classification is proficient with acceptable precision and recall (Table \ref{t:knn:metrics}). We saw a similar performance in \cite{Holwerda23} for the Hydra cluster. The expectation is that it will identify most of the perturbed galaxies in a sample from their \HI{} morphometrics this way with still some sizeable contamination, i.e. the sample will be mostly complete but somewhat contaminated. This would reduce the number of galaxies that would need to be inspected visually significantly.

\subsection{Galaxy Scaling Relations}

The galaxy scaling relations for these samples are rudimentary. Stellar mass and star-formation estimates are based on WISE photometry alone. For an in-depth discussion on the scaling relations for these galaxies, we refer the reader to Deg et al. (\textit{in preparation}). Our aim here was to determine if there were substantial differences between the perturbed and unperturbed marked samples and how this translated from training set to deployment set. In the case of the BTF relation, there is a marked difference in the scaling relation for the training set but this disappears for the deployment. In the stellar and \HI{} mass relation, there is a flatter relation for the perturbed subsample in both the training and deployment sample.  

If a scaling relation trend holds with the transition from training sample to deployment, does it build confidence in the observed effect? We note that between training and deployment sample, there is a difference in distance and thus resolution (the objects in the NGC 4808 and 4636 fields are closer). And that the training sample is still fairly small for training purposes. The fact that the BTF is functionally identical in NGC 5044 irrespective of class, could be attributed to these issues of distance and training sample size. But the persistence of flatter relation between stellar and \HI{} has more the appearance of an inherent difference between perturbed looking galaxies and unperturbed appearing ones. 

\section{Conclusions}
\label{s:conclusions}

We present a \HI{} morphometrics catalog for three WALLABY fields (centered on NGC4636, 4808 and 5044) observed as part of the early pilot WALLABY observations with ASKAP.

The NGC 5044 mosaic shows the greatest diversity of \HI{} morphologies and hence morphometrics. This is the richest catalog with a substantial number of unresolved detections well beyond the central object's distance. 

The NGC 4636 field has been studied in detail in \cite{Lin23d} using a mix of WALLABY and FAST data to identify those galaxies which are undergoing ram-pressure stripping and gravitational interactions such as tidal interactions or full mergers. All these are anecdotally already known to cause mild to severe changes in the \HI{} morphology. Using their flags for the three phenomena (ram-pressure stripping, tidal interaction, and mergers) as an all-encompassing ``perturbed'' label for both the NGC 4636 and the neighboring NGC 4808 field, we trained a nearest neighbours algorithm, using 2 neighbors and 6 features in the morphometrics space. The training sample is small but optimized like this performs reasonably well (Table \ref{t:knn:metrics}), minimizing variance as much as practical. Exactly which 6 features remains somewhat undetermined as kNN performs well with different combinations but the six in Table \ref{t:features} are our choice for this paper. 

The kNN classifier, even trained on a relatively small training sample performs well in the identification of the perturbed population, enough to identify the fraction of galaxies affected and identify individual galaxies with reasonable confidence. It is a marked improvement on a simple selection criterion based on one or two \HI{} morphometrics, both in stability of the identified fraction as accuracy. 

Applying this kNN classifier on the objects in the three WALLABY fields within a distance of 60 Mpc, we find ``perturbed'' populations in all three, mixed with the unperturbed population in most of the galaxy characteristics. The star-forming main sequence, to which most of these galaxies belong, is functionally the same for the perturbed and non-perturbed populations. 
The fact that both populations are well mixed-together in position  points to short time-scale effects, i.e. localized ones for the source of the perturbance, not throughout the field. 

We construct scaling relations for training and deployment samples using WISE W1 and W3 fluxes as proxies for stellar mass and star-formation rate and the SoFiA output. These are somewhat less precise as the scaling relations in Deg et al. (2024, \textit{in prep.}) but are only to be used to compare between the ``perturbed'' and ``unperturbed'' classes. 
The perturbed population does have a lower lower \HI{} mass compared to the stellar mass. The other scaling relations are indistinguishable from each other. We note that the Baryonic Tully-Fisher relation for the training sample shows a difference, while the deployment sample, the NGC 5044 field objects, does not, likely a result of the (still) low number statistics in the training sample. 

Our main result is a prediction for a study similar to that of \cite{Lin23d}: a list of candidate perturbed galaxies in the NGC 5044 mosaic. Once a similar study is conducted, a training sample for full deployment on all of WALLABY \HI{} morphometrics will be available.

\begin{acknowledgement}

This scientific work uses data obtained from Inyarrimanha Ilgari Bundara / the Murchison Radio-astronomy Observatory. We acknowledge the Wajarri Yamaji People as the Traditional Owners and native title holders of the Observatory site. CSIRO’s ASKAP radio telescope is part of the Australia Telescope National Facility (\url{https://ror.org/05qajvd42}). Operation of ASKAP is funded by the Australian Government with support from the National Collaborative Research Infrastructure Strategy. ASKAP uses the resources of the Pawsey Supercomputing Research Centre. Establishment of ASKAP, Inyarrimanha Ilgari Bundara, the CSIRO Murchison Radio-astronomy Observatory and the Pawsey Supercomputing Research Centre are initiatives of the Australian Government, with support from the Government of Western Australia and the Science and Industry Endowment Fund.

Parts of this research were supported by the Australian Research Council Centre of Excellence for All Sky Astrophysics in 3 Dimensions (ASTRO 3D), through project number CE170100013.

This research made use of Astropy, a community-developed core Python package for Astronomy \citep{Astropy-Collaboration13,Astropy-Collaboration18}.

\end{acknowledgement}

\paragraph{Funding Statement}

N.K.Y. acknowledges the China Postdoctoral Science Foundation (2022M723175, GZB20230766).

\paragraph{Data Availability Statement}

All ASKAP data products are publicly available in the CSIRO ASKAP Science Data Archive (CASDA\footnote{https://data.csiro.au/}).

ALL WALLABY PDR1 data is publicly available at WALLABY PDR1.
% https://doi.org/10.25919/xr39-jh63
% pls check if the above is correct
The kinematic modeling proto-pipeline is available at WKAPP code. The \HI{} morphometric catalogues are available with this paper. The specific morphometric and kNN analysis scripts are available upon request.

%\endnote in some journals will behave like \footnote; and \printendnotes will not output anything. 
% \printendnotes

%\printbibliography

% \printbibliography
% \bibliography{Bibliography.bib}

\appendix

% \section{The Star-formation and stellar mass plots with \HI{} Morphometrics}

\end{document}